\documentclass[10pt,onecolumn]{IEEEtran}
\usepackage{indentfirst}
\usepackage[dvips]{graphicx}
\usepackage{amsfonts}
\usepackage{multirow}
\usepackage{amsmath,amsthm,amssymb}
\usepackage{color,changepage}

\usepackage{CJK}
\usepackage{algorithm}
\usepackage{algorithmic}
\usepackage{bbm}
\usepackage{verbatim}
\usepackage{makecell}
\usepackage{subfigure}
\usepackage{mathrsfs}
\usepackage[numbers,sort&compress]{natbib}
\usepackage{arydshln}

\allowdisplaybreaks[4]

\definecolor{newcolor}{rgb}{0.5,0,1}

\newtheorem{Theorem}{Theorem}
\newtheorem{Corollary}{Corollary}
\newtheorem{Definition}{Definition}

\newtheorem{Lemma}{Lemma}

\theoremstyle{remark}
\newtheorem{Remark}{Remark}

\DeclareMathAlphabet{\mathpzc}{OT1}{pzc}{m}{it}

\begin{document}

\title{A new Capacity-Achieving Private Information Retrieval  Scheme with (Almost) Optimal File Length for Coded Servers}
\author{Jinbao Zhu, Qifa Yan, Chao Qi, and Xiaohu Tang, \IEEEmembership{Senior Member, IEEE}
\thanks{
J. Zhu and X. Tang are with the Information Security and National Computing Grid Laboratory, Southwest Jiaotong University, Chengdu, China (email: jinbaozhu@my.swjtu.edu.cn, xhutang@swjtu.edu.cn).

Q. Yan is with the  Department of Communications and Electronics, Telecom Paristech, 75634 Paris Cedex 13, France (email: qifa.yan@telecom-paristech.fr).

C. Qi is with the Signal Theory and Communications Department, Universidad Carlos III de Madrid, 28911 Leganes, Spain (email: chaoqi@hermod.tsc.uc3m.es).
}
}

\maketitle

\bibliographystyle{plain}
\thispagestyle{plain}\pagestyle{plain}

\begin{abstract}
In a distributed storage system,   private information retrieval (PIR)  guarantees  that  a user retrieves one file from the system without revealing any information about the identity of its interested file to any individual server.  In this paper, we investigate
$(N,K,M)$ coded sever model of PIR, where each of  $M$ files is distributed to the $N$ servers in the form of $(N,K)$ maximum distance separable (MDS) code for some $N>K$ and $M>1$.
As a result, we propose a new capacity-achieving $(N,K,M)$ coded linear PIR scheme such that it can be implemented with file length $\frac{K(N-K)}{\gcd(N,K)}$, which is much smaller than the previous best result $K\big(\frac{N}{\gcd(N,K)}\big)^{M-1}$. Notably, among all the capacity-achieving coded linear PIR schemes, we show that the file length is  optimal if $M>\big\lfloor \frac{K}{\gcd(N,K)}-\frac{K}{N-K}\big\rfloor+1$, and  within a multiplicative gap $\frac{K}{\gcd(N,K)}$ of a lower bound on the minimum file length otherwise.
\end{abstract}


\section{Introduction}
With the rapid development of open source systems, protecting user's privacy of retrieved information from public servers becomes a new challenge. This problem, referred to as private information retrieval (PIR) has direct applications in many scenarios such as cloud service, Internet of things, social networks, and so on. For examples,    PIR can ensure the privacy of the identity of the stocks downloaded by an investor from the stock market, and the privacy of activists against oppressive regimes while browsing some information deeded to be anti-regime. 

The PIR problem  was first introduced by Chor \textit{et al.} in \cite{Chor} and has drawn much attention from computer science community subsequently \cite{Gasarch,Cachin,Ostrovsky,Yekhanin}.
The classical formulation of PIR allows a user to query and download a file from  $N$ servers, each hosting the whole library of $M$ files, without revealing any information about the identity of the desired file  to any one of the servers. To retrieve a particular file, the user first sends a query string to each server. After receiving query, the server responds by sending an answer string to the user. Then the user decodes its requested file with these answers from the servers.

A native strategy  is to download all the files of the library no matter which file the user  needs. However, it is extremely inefficient in terms of the retrieval rate, 
which is defined as the number of bits that the user can privately retrieve  per  bit of downloaded data.
In the practical
systems, it is preferred to design  PIR schemes achieving as large as possible retrieval rate, especially achieving its  supremum, which is known as the capacity of the system.
In the initial work  \cite{Chor},   one bit length file was considered,  while the cost  was measured by the total lengths of all query and answer strings.
In fact, the corresponding  retrieval rate is $\Theta(1/M^{\frac{1}{N}})$.  Nevertheless, the Shannon theoretic formulation allows the length of files to be arbitrarily large, and therefore  the upload cost (total length of query strings) is  insignificant compared to the download cost (total length of answer strings), which is the more practical scenario. Under this assumption,
Shah \textit{et al.} proposed a scheme achieving the retrieval  rate $1-\frac{1}{N}$ \cite{N. B. Shah}.
In the later attractive work by Sun and Jafar \cite{Sun replicated}, the PIR capacity was finally characterized as $\left(1+\frac{1}{N}+\ldots+\frac{1}{N^{M-1}}\right)^{-1}$ for any $N$ and $M$.

The capacity-achieving PIR scheme in \cite{Sun replicated} requires the length of the files to be a multiple of $N^M$, which increases to infinity exponentially with number of files $M$. Although large file length can contribute to improve retrieval rate, it  also arouses complexity increase in practical implementations. The problem of decreasing the file length, also known as sub-packetization in the literature, has been noted in many applications, for example coded caching \cite{coded caching1,coded caching2,coded caching3}, while it becomes the bottleneck for practical application of the PIR scheme in \cite{Sun replicated}. The file length of capacity-achieving PIR schemes was reduced to $N^{M-1}$ in \cite{Sun optimal} subsequently. Most notably, the optimal length was  proved to be $N-1$ in \cite{Tian and Sun} very recently, which turns out to be independent of $M$ and much smaller than previous schemes.

It should be noted that all the aforementioned results are obtained in the classical setup, which requires each sever to store all the files completely, i.e., a replication code is used to store the contents.
Though repetition coding  can offer the highest capability against sever failures
and simplify the  implementation of PIR schemes as well, it incurs pretty large storage cost. This impels the use of erasure coding techniques, especially  the maximum distance separable (MDS) code that achieves the optimal redundancy-reliability tradeoff \cite{A. G. Dimakis}.  In \cite{EI Rouayheb},
Tajeddine and Rouayheb considered a MDS coded  PIR model that each file is stored across $N$ servers through an $(N,K)$ MDS code, and  proposed a PIR scheme achieving the retrieval rate $1-\frac{K}{N}$, where the servers are referred to as \emph{coded servers}. This model allows each server to store only $\frac{1}{K}$ of each file in the library and degrades to the classical setup when the code used is an $(N,1)$ repetition code. The PIR capacity of MDS coded server system was found by  Banawan and Ulukus in \cite{Ulukus} to be $\big(1+\frac{K}{N}+...+\big(\frac{K}{N}\big)^{M-1}\big)^{-1}$ where the capacity-achieving scheme requires that, each file has to be at least length $KN^M$. The file length problem again shows up in this more generalized model. In \cite{Zhang}, the authors proposed a capacity-achieving scheme, whose file length is decreased to $K\big(\frac{N}{\gcd(N,K)}\big)^{M-1}$, but still increases  exponentially with number of files $M$.

For the coded servers, the exciting result  \cite{Tian and Sun}  in the classical setup motivates us to explore the problem: Does there exist a capacity-achieving coded scheme under this more general setup such that, the length of files  is independent of $M$? In this paper, we give an affirmative answer.
The previous results in \cite{Ulukus} and \cite{Zhang} indicates that, \emph{linear} PIR scheme is sufficient to achieve the capacity even in the case of coded servers, i.e., the answers can be restricted to linear functions of the stored contents.
So, we only focus on  capacity-achieving coded linear PIR schemes due to their  simplicity. Moreover, we discard the trivial cases of $N=K$ or $M=1$, since the user can download all contents stored by the servers to guarantee privacy in  the former case and  essentially no file can be protected in the latter case. For that reason, the objective of  this paper is to design capacity-achieving coded linear PIR scheme with practical file length for the coded server system in the cases of $N>K$ and $M>1$.
The contributions of this paper are two folds:
\begin{enumerate}
  \item We propose a new  capacity-achieving coded linear PIR scheme with file length $L=\frac{K(N-K)}{\gcd(N,K)}$.
The new scheme has obvious advantage with respect to  file length over the state-of-the-art in \cite{Ulukus} and \cite{Zhang}. For clarity, we compare them in Table \ref{tab:comparative}.
\begin{table*}[htbp]
\extrarowheight=4pt
\centering
\caption{Comparison of  Capacity-Achieving PIR Schemes for Coded Servers}
  \begin{tabular}{|c|c|c|c|}
  \hline
  Capacity-Achieving PIR Schemes & Banawan-Ulukus scheme & Xu-Zhang scheme &  New scheme \\
  \hline
 Minimum File Length & $KN^M$ & $K\left(\frac{N}{\gcd(N,K)}\right)^{M-1}$ & $\frac{K(N-K)}{\gcd{(N,K)}}$ \\
  \hline
  \end{tabular}
  \label{tab:comparative}
\end{table*}
  \item We show that  the file length $L$ is optimal among all the capacity-achieving $(N,K,M)$ coded  linear PIR schemes when
$M>\big\lfloor \frac{K}{\gcd(N,K)}-\frac{K}{N-K}\big\rfloor+1$. In the other case, the file length $L$ is within a multiplicative gap $\frac{K}{\gcd(N,K)}$ in contrast  to  its
lower bound.

\end{enumerate}

The rest of this paper is organized as follows. In Section \ref{system model},  the MDS coded server model  is formally described and linear PIR scheme is characterized. In Section \ref{new coded scheme}, a new capacity-achieving coded linear  PIR scheme is proposed. In Section \ref{sec:achievability}, the achievability of the new coded scheme is analyzed. In Section \ref{the converse bound}, the optimality of the file length is established. Finally, the paper is concluded  in Section \ref{conclusion}.

Throughout this paper, the following notations are used.
\begin{itemize}
\item  For two non-negative integers  $n$ and $m$ with $n<m$, define  $[n:m)$ as the set $\{n,n+1,\ldots, m-1\}$;
\item For a finite set $S$, $|S|$ denotes its cardinality;
\item Denote $A_{0:m-1}$ an ordered set $\{A_0,\ldots,A_{m-1}\}$, and define $A_{\Gamma}$ as  $\{A_{\gamma_0},\ldots,A_{\gamma_{k-1}}\}$ for any  indices set $\Gamma=\{\gamma_0,\ldots,\gamma_{k-1}\}$\\$\subseteq[0:m)$ with $\gamma_0<\ldots<\gamma_{k-1}$;
\item For a matrix $\mathcal{Q}$, define $\mathcal{Q}(i,:)$, $\mathcal{Q}(:,j)$ and $\mathcal{Q}(i,j)$ as its $i$-th row vector, its $j$-th column vector and the element in  row $i$ and column $j$, respectively.
\end{itemize}

\section{System Model}\label{system model}
Consider a distributed storage system that stores $M$  encoded files  $\mathcal{W}_0,\mathcal{W}_1,\ldots,\mathcal{W}_{M-1}$ across $N$ servers  by using a fixed $(N,K)$ MDS code over $\mathbb{F}_p$ for  a  prime power $p$. The $M$ files are independent of each other and each file $\mathcal{W}_i$ is of the form
 \begin{equation}
\mathcal{W}_{i} =
\left(
  \begin{array}{ccc}
    \mathcal{W}_i (0,0) & \ldots & \mathcal{W}_i (0,N-1) \\
    \vdots & \ddots & \vdots \\
    \mathcal{W}_i (\lambda -1,0) & \ldots &\mathcal{W}_i(\lambda -1,N-1)
\end{array}
\right),~i\in[0:M),\label{encoded symbols}
\end{equation}
where the row vectors $\mathcal{W}_i(0,:),\mathcal{W}_i(1,:),\ldots,\mathcal{W}_i(\lambda-1,:)$ are $\lambda$   independent codewords of the $(N,K)$ MDS code.
We refer to the quantity
\begin{IEEEeqnarray}{c}
L\triangleq K\cdot\lambda\notag
\end{IEEEeqnarray}
as \emph{file length}, since each file can be equivalently represented by a \emph{source file} consisting of  $\lambda$ vectors of length $K$ over $\mathbb{F}_p$.
As a consequence,
\begin{IEEEeqnarray}{rCl}
H( \mathcal{W}_0)=\ldots=H( \mathcal{W}_{M-1})&=&L,\label{Eqn_W_Size}\\
H( \mathcal{W}_0,\ldots, \mathcal{W}_{M-1})&=&\sum_{i=0}^{M-1}H(\mathcal{W}_i),
\end{IEEEeqnarray}
where the entropy function $H(\cdot)$ is measured with logarithm  $p$.

At the $t$-th server, the stored contents $\mathbf{y}_t\in\mathbb{F}_{p}^{M\lambda }$ are the concatenation of the $t$-th column in all the encoded files $\mathcal{W}_{0:M-1}$ and are given by
 \begin{equation}
\mathbf{y}_{t}=
\left(
  \begin{array}{c}
    \mathcal{W}_{0}(:,t) \\
    \mathcal{W}_{1}(:,t) \\
    \vdots \\
    \mathcal{W}_{M-1}(:,t) \\
  \end{array}
\right),~t\in[0:N). \notag
\end{equation}

Due to the property of the base $(N,K)$ MDS code,  the storage system can reconstruct all the files by connecting to any $K$ servers to tolerate  up to $N-K$ server failures. Then, for any set $\Gamma\subseteq[0:N)$ with  $|\Gamma|\geq K$,
\begin{IEEEeqnarray}{c}
H(\mathbf{y}_{\overline{\Gamma}}|\mathbf{y}_{\Gamma})=0, \notag
\end{IEEEeqnarray}
where $\overline{\Gamma}\triangleq[0:N)\backslash\Gamma$.

A user privately generates $\theta\in[0:M)$ and wishes to retrieve $\mathcal{W}_{\theta}$ from the storage system\footnote{Note that, the user can retrieve the encoded file $\mathcal{W}_\theta$ if and only if it can retrieve  $\theta$-th source file.}  while ensuring that any one of the $N$ servers will have no information about $\theta$, by means of a \emph{private information retrieval}
(\emph{PIR}) scheme consisting of the following phases:
\begin{enumerate}
  \item \emph{Query Phase:} The user randomly generates a set of queries $\mathcal{Q}_{0:N-1}^{[\theta]}$ according to some distribution over $\mho^N$, and sends $\mathcal{Q}_{t}^{[\theta]}$ to the $t$-th sever, where the set $\mho$ will be referred to as \emph{query space}. Indeed, the queries are generated independently of file realizations, i.e.,
      \begin{IEEEeqnarray}{c}\label{query independence}
      I(\mathcal{Q}_{0:N-1}^{[\theta]};\mathcal{W}_{0:M-1})=0.
      \end{IEEEeqnarray}
  \item \emph{Answer Phase:} Upon receiving the query $\mathcal{Q}_{t}^{[\theta]}$, the $t$-th server responds with an answer $\mathbf{A}_{t}^{[\theta]}$, which is a deterministic function of the received query and the stored contents at  server $t$. Thus, by  the data-processing theorem,
      \begin{IEEEeqnarray}{c}\label{answer function}
        H(\mathbf{A}_{t}^{[\theta]}|\mathcal{Q}_{t}^{[\theta]},\mathbf{y}_t)=H(\mathbf{A}_{t}^{[\theta]}|\mathcal{Q}_{t}^{[\theta]},\mathcal{W}_{0:M-1})=0,~t\in[0:N).
      \end{IEEEeqnarray}
  \item \emph{Decoding Phase:} The user must correctly decode the desired file $ \mathcal{W}_{\theta}$ from answers $\mathbf{A}_{0:N-1}^{[\theta]}$.

\end{enumerate}
As a PIR scheme, it has to satisfy
\begin{itemize}
  \item\textbf{Correctness:} The file $ \mathcal{W}_{\theta}$ can be completely disclosed by the queries and answers, i.e., 
     \begin{IEEEeqnarray}{c}\label{correctness constrain}
        H( \mathcal{W}_{\theta}\;|\;\mathbf{A}_{0:N-1}^{[\theta]},\;\mathcal{Q}_{0:N-1}^{[\theta]})=0,~\theta\in[0:M).
      \end{IEEEeqnarray}

 \item\textbf{Privacy:} 
Each server $t\in[0:N)$ should learn nothing about  the index $\theta$, i.e.,
  \begin{IEEEeqnarray}{c}
 I(\mathcal{Q}_t^{[\theta]},\mathbf{A}_{t}^{[\theta]},\mathbf{y}_t;\theta)=0,
  \label{infor-theoretic privacy}
  \end{IEEEeqnarray}
  where $(\mathcal{Q}_{t}^{[\theta]},\mathbf{A}_{t}^{[\theta]},\mathbf{y}_{t})$ is all the information owned by the $t$-th server. Equivalently, its distribution  is independent of the realization of $\theta$, i.e.,
 \begin{IEEEeqnarray}{c}
 (\mathcal{Q}_t^{[\theta]},\mathbf{A}_{t}^{[\theta]},\mathbf{y}_t)\sim (\mathcal{Q}_t^{[\theta']},\mathbf{A}_{t}^{[\theta']},\mathbf{y}_t), \quad\forall~\theta,\theta'\in[0:M),\; t\in[0:N), \label{idential distributed}
 \end{IEEEeqnarray}
 where $X\sim Y$ indicates that $X$ and $Y$ are identically distributed.
\end{itemize}
We call the above PIR  an $(N,K,M)$ coded PIR scheme.

The \emph{rate} of a coded PIR scheme, denoted by $R$ is defined as
  \begin{IEEEeqnarray}{c}\label{Moded rate}
                R=\frac{L}{D},
            \end{IEEEeqnarray}
            where $D$ is download cost (the total length of the answers) from the $N$ servers  averaged  over random query realizations.

 Notice that the $R$ and $D$ are independent of $\theta$ by the privacy constraint in \eqref{idential distributed}.

\begin{Definition}[The Capacity of $(N,K,M)$ Coded PIR] Given any $(N,K,M)$, a rate $R$ is said to be achievable if there exists a coded PIR scheme with rate greater than or equal to $R$. The capacity of the coded PIR, denoted by $C$, is defined as the supremum over all the achievable rates.
\end{Definition}
The capacity of the $(N,K,M)$  coded PIR scheme has been determined in \cite{Ulukus}  as
            \begin{IEEEeqnarray}{c}
                C=\left(1+\frac{K}{N}+\left(\frac{K}{N}\right)^2+\ldots+\left(\frac{K}{N}\right)^{M-1}\right)^{-1}.\label{the capacity}
            \end{IEEEeqnarray}
The works in \cite{Ulukus} and \cite{Zhang}  indicate that coded linear PIR schemes suffice to achieve the  capacity.

\begin{Definition}[Coded Linear PIR Scheme] For an $(N,K,M)$ coded PIR scheme, let $\ell_{t}$ be the \emph{answer length} of the received query $\mathcal{Q}_{t}^{[\theta]}$ at server $t$.
It is said to be an $(N,K,M)$ coded linear PIR scheme
 if  the answers $\mathbf{A}_t^{[\theta]}(t\in[0:N))$ are formed by
 \begin{IEEEeqnarray}{c}\label{linear answer function}
      \mathbf{A}_{t}^{[\theta]}=\mathcal{A}_{t,[0:M)}^{[\theta]}\mathbf{y}_{t},~t\in[0:N),
      \end{IEEEeqnarray}
where $\mathcal{A}_{t,[0:M)}^{[\theta]}=( \mathcal{A}_{t,0}^{[\theta]},\ldots,\mathcal{A}_{t,M-1}^{[\theta]})
$ is an answer matrix of order $\ell_t\times M\lambda $ over $\mathbb{F}_p$  only determined  by the  query $\mathcal{Q}_{t}^{[\theta]}$, and
$\mathcal{A}_{t,i}^{[\theta]}(i\in[0:M))$ is a sub-matrix of order $\ell_t\times\lambda$. In particular, we call
 \begin{equation*}
 \mathcal{A}_{t,[0:M)\backslash\{\theta\}}^{[\theta]}=\left( \mathcal{A}_{t,0}^{[\theta]},\ldots, \mathcal{A}_{t,\theta-1}^{[\theta]},\mathcal{A}_{t,\theta+1}^{[\theta]},\ldots,\mathcal{A}_{t,M-1}^{[\theta]}\right)
 \end{equation*}
 the answer-interference matrix.
\end{Definition}

  Hence, the download cost $D$ of a linear PIR scheme can be calculated as
              \begin{IEEEeqnarray}{c}
            D=\mathbb{E}\left[\sum_{t=0}^{N-1}\ell_t\right]. \notag
            \end{IEEEeqnarray}

The objective of this paper is to design $(N,K,M)$ coded linear PIR schemes which simultaneously achieve the PIR capacity and possess the practical file length in the non-trivial cases of $N>K$ and $M>1$.
Obviously, for a   linear PIR scheme,  all the queries and answers can be viewed as matrices. Thus, from now on, for the sake of unification we use cursive capital letters to denote random matrices such as $\mathcal{Q},\mathcal{A}$,
and $\widetilde{\mathcal{Q}}$ and $\widetilde{\mathcal{A}}$ as their realizations. Besides, we denote the column vectors by the bold letters, e.g. $\mathbf{A},\mathbf{q}$.

\section{A New Capacity-Achieving Coded Linear PIR Scheme}\label{new coded scheme}

In this section, we propose a new capacity-achieving coded linear PIR scheme with the file length $\frac{K(N-K)}{\gcd(N,K)}$ for $N>K$ and $M>1$, and then give an illustrative example.

\subsection{A New Coded Linear PIR Scheme}

Define
\begin{IEEEeqnarray}{rCl}
n\triangleq\frac{N}{\gcd(N,K)}, \notag\\
k\triangleq\frac{K}{\gcd(N,K)}. \notag
\end{IEEEeqnarray}

Consider the case that each file has length $L=K\cdot\lambda $, where
\begin{IEEEeqnarray}{c}
\lambda =n-k.\label{subpacketization}
\end{IEEEeqnarray}
That is to say, each column vector of all encoded files in \eqref{encoded symbols} is  of length $\lambda =n-k$. For easy of exposition, we append $k$ dummy zeros to  the vectors of each file stored at each server, i.e., the dummy expanded vectors at server $t$  are
\begin{IEEEeqnarray}{c}
\mathcal{W}_{i}(:,t)\triangleq \big(\mathcal{W}_{i}(0,t),\ldots,\mathcal{W}_{i}(n-k -1,t),\underbrace{0,\ldots,0}_{k}\big)^\top,~i\in[0:M),\label{supply zero}
\end{IEEEeqnarray}
where $\top$ denotes the transpose operator.

Denote $\Omega$ the set of vectors in $[0:n)^k$ with distinct entries in all coordinates, i.e.,
\begin{IEEEeqnarray}{rCl}
\Omega&=&\left\{\mathbf{q}=(\mathbf{q}(0),\ldots,\mathbf{q}(k-1))^\top\in[0:n)^k:\mathbf{q}(i)\neq\mathbf{q}(j),~\forall~i,j\in[0:k),\; i\neq j\right\}. \notag
\end{IEEEeqnarray}

We are now ready to present the new coded linear PIR scheme.
\begin{enumerate}
  \item \textbf{Query Phase: }Assume that  file $ \mathcal{W}_{\theta}$ is requested by the user.
The user first generates a $k\times M$ random matrix
\begin{IEEEeqnarray}{c}
\mathcal{Q}=\left(\mathbf{q}_0,\ldots,\mathbf{q}_{\theta-1},\mathbf{q}_{\theta},\mathbf{q}_{\theta+1},\ldots,\mathbf{q}_{M-1}\right),\label{eqn:Q_theta}
\end{IEEEeqnarray}
where each column vector $\mathbf{q}_i$ is drawn independently and uniformly from $\Omega$ for all $0\leq i<M$. 

Next, the user constructs the queries based on the matrix $\mathcal{Q}$ as
        \begin{eqnarray}\label{query formulation}
        \mathcal{Q}_{t}^{[\theta]}=\left(\mathbf{q}_0,\ldots,\mathbf{q}_{\theta-1},\left(\mathbf{q}_{\theta}+t\right)_n,\mathbf{q}_{\theta+1},\ldots,\mathbf{q}_{M-1}\right),~t\in[0:N),
        \end{eqnarray}
        where $\mathbf{q}_{\theta}+t$ denotes the vector that each element in the column vector $\mathbf{q}_{\theta}$ is added $t$, i.e.,
        $\mathbf{q}_{\theta}+t=(\mathbf{q}_{\theta}(0)+ t,\ldots, \mathbf{q}_{\theta}(k-1)+t)^\top$ and $(\cdot)_n$ denotes the element-wise modulo $n$ operation. Notice that $k\times M$ matrix $\mathcal{Q}_t^{[\theta]}(0\leq t<N)$ is almost the same as $\mathcal{Q}$ with the exception of the $\theta$-th column, and $\mathcal{Q}_{0:N-1}^{[\theta]}$ satisfies \eqref{query independence} since each $\mathcal{Q}_{t}^{[\theta]}$ only depends on $\mathcal{Q},t,\theta$, which are all independent of $\mathcal{W}_{0:M-1}$.

 \item \textbf{Answer Phase: } The servers respond in $k$ rounds, each round indexed by an integer $s\in[0:k)$. In particular, in the round $s$, server $t$ sends
      \begin{IEEEeqnarray}{c}
      \mathbf{A}_{t}^{[\theta]}(s)=\left\{\begin{array}{ll}
                                \textnormal{NULL},   &\text{if}~\mathcal{Q}_{t}^{[\theta]}(s,:)\in[n-k:n)^M  \\
                                 \sum\limits_{i\in[0:M)}\mathcal{W}_{i}\big(\mathcal{Q}_t^{[\theta]}(s,i),t\big),&\text{else}
                                \end{array}
      \right.\label{server answer}
      \end{IEEEeqnarray}
      to the user, where the value $\textnormal{NULL}$ indicates that the server keeps silence, and the additions are operated on the finite field $\mathbb{F}_p$.
 \item \textbf{Decoding Phase: } We defer the reconstruction procedure in Section \ref{correct lemma}.
\end{enumerate}

It is easy to see from \eqref{server answer} that there exists answer matrices  $\mathcal{A}_{t,[0:M)}^{[\theta]}$ satisfying \eqref{linear answer function}. That is, the proposed scheme is actually a linear PIR scheme.

For the above scheme, we have the following  main results of this paper.
\begin{Theorem}\label{thm:main}
Given any $(N,K,M)$ with $N>K$ and $M>1$, there exists a capacity-achieving coded linear PIR scheme with file length $L=\frac{K(N-K)}{\gcd(N,K)}$. The file length $L$ is optimal among all the capacity-achieving $(N,K,M)$ coded  linear PIR schemes if
$M>\big\lfloor \frac{K}{\gcd(N,K)}-\frac{K}{N-K}\big\rfloor+1$. Otherwise, the file length $L$ is within a multiplicative gap $\frac{K}{\gcd(N,K)}$ compared to a lower bound on the minimum file length of capacity-achieving
$(N,K,M)$ coded  linear PIR schemes.
\end{Theorem}
\begin{IEEEproof}
The proofs of correctness, privacy, and performance of the new scheme are given in Section \ref{sec:achievability}-A, B, and C respectively. The optimality and multiplicative gap of the file length are shown in Theorems \ref{The_general bound} and \ref{The_tight bound}.
\end{IEEEproof}

\subsection{An Illustrative Example}

In this subsection, we illustrate an example of $(N,K,M)=(5,3,3)$ coded linear PIR scheme. According to \eqref{subpacketization},
$\lambda =2$ and $L=K\cdot\lambda=6$.  The three encoded files $ \mathcal{W}_0, \mathcal{W}_1, \mathcal{W}_2$ are respectively denoted as
\begin{IEEEeqnarray}{rCl}
\mathcal{W}_i&=&
\left(
  \begin{array}{ccccc}
    \mathcal{W}_i(0,0) & \mathcal{W}_i(0,1) & \mathcal{W}_i(0,2) & \mathcal{W}_i(0,3) & \mathcal{W}_i(0,4) \\
    \mathcal{W}_i(1,0) & \mathcal{W}_i(1,1) & \mathcal{W}_i(1,2) & \mathcal{W}_i(1,3) & \mathcal{W}_i(1,4)
 \end{array}
\right), ~i=0,1,2, \notag
\end{IEEEeqnarray}
where each row vector forms a $(5,3)$ MDS codeword. Then, they are stored across $5$ servers.  To better understand the corresponding relationship of the stored contents, the query matrices and the answers, the former  at each sever are arranged as a matrix as shown in Fig. \ref{fig:storage}.
Notably, the symbols $0$ in the boxes with dotted lines are  the appended dummy zeros and not stored at all.

\begin{figure*}[htbp]
\centering
\begin{tabular}{;{2pt/2pt}c;{2pt/2pt}c;{2pt/2pt}c;{2pt/2pt}}
  \multicolumn{3}{@{}c@{}}{Server 0}  \\
  \hline
  \multicolumn{1}{|c|}{$\mathcal{W}_0(0,0)$} &  \multicolumn{1}{c|}{$\mathcal{W}_1(0,0)$} & \multicolumn{1}{c|}{$\mathcal{W}_2(0,0)$}  \\
  \hline
  \multicolumn{1}{|c|}{$\mathcal{W}_0(1,0)$} &  \multicolumn{1}{c|}{$\mathcal{W}_1(1,0)$}  & \multicolumn{1}{c|}{$\mathcal{W}_2(1,0)$}  \\
  \hdashline[2pt/2pt]
  $0$ & $0$ & $0$  \\
  \hdashline[2pt/2pt]
  $0$ & $0$ & $0$  \\
  \hdashline[2pt/2pt]
  $0$ & $0$ & $0$ \\
  \hdashline[2pt/2pt]
\end{tabular}
\begin{tabular}{;{2pt/2pt}c;{2pt/2pt}c;{2pt/2pt}c;{2pt/2pt}}
  \multicolumn{3}{@{}c@{}}{Server 1}  \\
  \hline
  \multicolumn{1}{|c|}{$\mathcal{W}_0(0,1)$} &  \multicolumn{1}{c|}{$\mathcal{W}_1(0,1)$} & \multicolumn{1}{c|}{$\mathcal{W}_2(0,1)$}  \\
  \hline
  \multicolumn{1}{|c|}{$\mathcal{W}_0(1,1)$} &  \multicolumn{1}{c|}{$\mathcal{W}_1(1,1)$}  & \multicolumn{1}{c|}{$\mathcal{W}_2(1,1)$}  \\
  \hdashline[2pt/2pt]
  $0$ & $0$ & $0$  \\
  \hdashline[2pt/2pt]
  $0$ & $0$ & $0$  \\
  \hdashline[2pt/2pt]
  $0$ & $0$ & $0$ \\
  \hdashline[2pt/2pt]
\end{tabular}
\begin{tabular}{;{2pt/2pt}c;{2pt/2pt}c;{2pt/2pt}c;{2pt/2pt}}
  \multicolumn{3}{@{}c@{}}{Server 2}  \\
  \hline
  \multicolumn{1}{|c|}{$\mathcal{W}_0(0,2)$} &  \multicolumn{1}{c|}{$\mathcal{W}_1(0,2)$} & \multicolumn{1}{c|}{$\mathcal{W}_2(0,2)$}  \\
  \hline
  \multicolumn{1}{|c|}{$\mathcal{W}_0(1,2)$} &  \multicolumn{1}{c|}{$\mathcal{W}_1(1,2)$}  & \multicolumn{1}{c|}{$\mathcal{W}_2(1,2)$}  \\
  \hdashline[2pt/2pt]
  $0$ & $0$ & $0$  \\
  \hdashline[2pt/2pt]
  $0$ & $0$ & $0$  \\
  \hdashline[2pt/2pt]
  $0$ & $0$ & $0$ \\
  \hdashline[2pt/2pt]
\end{tabular}
\begin{tabular}{;{2pt/2pt}c;{2pt/2pt}c;{2pt/2pt}c;{2pt/2pt}}
  \multicolumn{3}{@{}c@{}}{Server 3}  \\
  \hline
  \multicolumn{1}{|c|}{$\mathcal{W}_0(0,3)$} &  \multicolumn{1}{c|}{$\mathcal{W}_1(0,3)$} & \multicolumn{1}{c|}{$\mathcal{W}_2(0,3)$}  \\
  \hline
  \multicolumn{1}{|c|}{$\mathcal{W}_0(1,3)$} &  \multicolumn{1}{c|}{$\mathcal{W}_1(1,3)$}  & \multicolumn{1}{c|}{$\mathcal{W}_2(1,3)$}  \\
  \hdashline[2pt/2pt]
  $0$ & $0$ & $0$  \\
  \hdashline[2pt/2pt]
  $0$ & $0$ & $0$  \\
  \hdashline[2pt/2pt]
  $0$ & $0$ & $0$ \\
  \hdashline[2pt/2pt]
\end{tabular}
\begin{tabular}{;{2pt/2pt}c;{2pt/2pt}c;{2pt/2pt}c;{2pt/2pt}}
  \multicolumn{3}{@{}c@{}}{Server 4}  \\
  \hline
  \multicolumn{1}{|c|}{$\mathcal{W}_0(0,4)$} &  \multicolumn{1}{c|}{$\mathcal{W}_1(0,4)$} & \multicolumn{1}{c|}{$\mathcal{W}_2(0,4)$}  \\
  \hline
  \multicolumn{1}{|c|}{$\mathcal{W}_0(1,4)$} &  \multicolumn{1}{c|}{$\mathcal{W}_1(1,4)$}  & \multicolumn{1}{c|}{$\mathcal{W}_2(1,4)$}  \\
  \hdashline[2pt/2pt]
  $0$ & $0$ & $0$  \\
  \hdashline[2pt/2pt]
  $0$ & $0$ & $0$  \\
  \hdashline[2pt/2pt]
  $0$ & $0$ & $0$ \\
  \hdashline[2pt/2pt]
\end{tabular}
\caption{The stored contents at each server.}\label{fig:storage}
\label{Fig:example}
\end{figure*}

Let
\begin{IEEEeqnarray}{rCl}
\Omega&=&\left\{\mathbf{q}\in[0:5)^3:\mathbf{q}(0)\neq\mathbf{q}(1),\mathbf{q}(0)\neq\mathbf{q}(2),\mathbf{q}(1)\neq\mathbf{q}(2)\right\}. \notag
\end{IEEEeqnarray}
In the query phase, the user  generates a $3\times3$ matrix $\mathcal{Q}$ with each column chosen independently and uniformly from $\Omega$, for example,
 \begin{IEEEeqnarray}{c}
\mathcal{Q}=\left(\mathbf{q}_0,\mathbf{q}_1,\mathbf{q}_{2}\right)=
\left(
  \begin{array}{ccc}
    3 & 4 & 3 \\
    0 & 1 & 0 \\
    1 & 0 & 4 \\
  \end{array}
\right). \notag
\end{IEEEeqnarray}
Assume that  $\mathcal{W}_0$ is the desired file.   Then,  the user sends the following query matrices to the severs:
\begin{IEEEeqnarray}{c}
\begin{tabular}{@{}c@{}c@{}c@{}c@{}c@{}}
Server 0 & Server 1 & Server 2 & Server 3 & Server 4 \\
$\left(
     \begin{array}{ccc}
       3 & 4 & 3 \\
       0 & 1 & 0 \\
       1 & 0 & 4 \\
     \end{array}
   \right)$
&
$\left(
     \begin{array}{ccc}
       4 & 4 & 3 \\
       1 & 1 & 0 \\
       2 & 0 & 4 \\
     \end{array}
   \right)$
&
$\left(
     \begin{array}{ccc}
       0 & 4 & 3 \\
       2 & 1 & 0 \\
       3 & 0 & 4 \\
     \end{array}
   \right)$
&
$\left(
     \begin{array}{ccc}
       1 & 4 & 3 \\
       3 & 1 & 0 \\
       4 & 0 & 4 \\
     \end{array}
   \right)$
&
$\left(
     \begin{array}{ccc}
       2 & 4 & 3 \\
       4 & 1 & 0 \\
       0 & 0 & 4 \\
     \end{array}
   \right)$ \\
\end{tabular}\notag
\end{IEEEeqnarray}

Upon receiving the queries, the servers  answer in $k=3$ rounds, which are referred to as round R0, R1 and R2 as shown in Fig. \ref{fig:answer}.
\begin{figure*}[htbp]
\centering
\begin{tabular}{@{}c@{\;\;\;}c@{\quad}c@{\quad}c@{}}
 & Server 0 & Server 1 & Server 2 \\
R0 & NULL & NULL & $\mathcal{W}_0(0,2)+0+0$ \\
R1 & $\mathcal{W}_{0}(0,0)+\mathcal{W}_{1}(1,0)+\mathcal{W}_{2}(0,0)$ & $\mathcal{W}_{0}(1,1)+\mathcal{W}_{1}(1,1)+\mathcal{W}_{2}(0,1)$ & $0+\mathcal{W}_{1}(1,2)+\mathcal{W}_{2}(0,2)$  \\
R2 & $\mathcal{W}_{0}(1,0)+\mathcal{W}_{1}(0,0)+0$ & $0+\mathcal{W}_{1}(0,1)+0$  & $0+\mathcal{W}_{1}(0,2)+0$ \\
 & Server 3 & Server 4 \\
R0 & $\mathcal{W}_{0}(1,3)+0+0$ & NULL\\
R1 & $0+\mathcal{W}_{1}(1,3)+\mathcal{W}_{2}(0,3)$ & $0+\mathcal{W}_{1}(1,4)+\mathcal{W}_{2}(0,4)$ \\
R2 & $0+\mathcal{W}_{1}(0,3)+0$ & $\mathcal{W}_{0}(0,4)+\mathcal{W}_{1}(0,4)+0$ \\
\end{tabular}
\caption{The answers at different servers.}\label{fig:answer}
\end{figure*}

In decoding phase,  the user decodes two symbols of  $ \mathcal{W}_0$ from the answers of each round:
\begin{enumerate}
    \item In R0, get $\mathcal{W}_{0}(0,2)$ and $\mathcal{W}_{0}(1,3)$ directly;
    \item In R1, decode the two interference symbols $\mathcal{W}_{1}(1,0)+\mathcal{W}_{2}(0,0),\mathcal{W}_{1}(1,1)+\mathcal{W}_{2}(0,1)$ from $\mathcal{W}_{1}(1,2)+\mathcal{W}_{2}(0,2),\mathcal{W}_{1}(1,3)+\mathcal{W}_{2}(0,3),\mathcal{W}_{1}(1,4)+\mathcal{W}_{2}(0,4)$ by MDS property and then get $\mathcal{W}_{0}(0,0)$ and $\mathcal{W}_{0}(1,1)$;
    \item In R2, similarly to R1, get $\mathcal{W}_{0}(1,0)$ and $\mathcal{W}_{0}(0,4)$.
\end{enumerate}
Then, the user is able to recover  $\mathcal{W}_0(0,:)$ and $\mathcal{W}_0(1,:)$ and thus the file $\mathcal{W}_0$ by means of MDS property again.

Privacy is guaranteed, since each column of the query matrix received by each server is uniformly and independently distributed on $\Omega$, regardless of the file being requested.

In this realization,  the download cost is $12$. Among all the realizations, since  the probability of each signal being NULL is $\big(\frac{3}{5}\big)^3$, and there are $3\times 5$ signals sent in total, the average download cost is
\begin{IEEEeqnarray}{c}
D=3\times 5\times \left(1-\left(\frac{3}{5}\right)^3\right)=\frac{294}{25}. \notag
\end{IEEEeqnarray}
The file length is $L=3\times 2=6$,
thus the retrieve rate is
\begin{IEEEeqnarray}{c}
R=\frac{L}{D}=\frac{25}{49}, \notag
\end{IEEEeqnarray}
which achieves the capacity of  $(5,3,3)$ coded PIR scheme in \eqref{the capacity}.

\section{The Achievability of New Coded Scheme}\label{sec:achievability}

In this section, we analyse the correctness, privacy, and performance of the scheme.
Before that, we need three obvious facts from the matrix $\mathcal{Q}$ and $\mathcal{Q}_{t}^{[\theta]}$ in \eqref{eqn:Q_theta} and \eqref{query formulation}:
\begin{enumerate}
   \item [\textbf{F1:}] For each $s\in[0:k)$,  $(\mathcal{Q}(s,\theta)+t)_n$ takes each value in the set $[0:n)$  exactly $d=\gcd(N,K)$ times, with  $t$ ranging over the set $[0:N)$;
   \item [\textbf{F2:}] For $s_1\ne s_2\in[0:k)$,  $\mathcal{Q}(s_1,\theta)\neq \mathcal{Q}(s_2,\theta)$;
   \item [\textbf{F3:}] For any given $s\in [0:k)$, $\{\mathcal{Q}(s,j):j\in[0:M)\}$ are independent and uniform in $[0:n)$, and so does $\{\mathcal{Q}^{[\theta]}_t(s,j):j\in[0:M)\}$ for all $\theta\in [0:M)$.
\end{enumerate}

\subsection{Proof of Correctness}\label{correct lemma}
For any realizations of random variables $\theta$ and $\mathcal{Q}$ in the new coded linear PIR scheme, the file $ \mathcal{W}_{\theta}$ can be reconstructed from  $\mathbf{A}_{0:N-1}^{[\theta]}$ and $\mathcal{Q}_{0:N-1}^{[\theta]}$. We now describe the decoding processes in detail.

First of all, by \eqref{supply zero} and \eqref{server answer},   no matter the answer $\mathbf{A}_{t}^{[\theta]}(s)$ of the $s$-th round of the received query matrix $\mathcal{Q}_t^{[\theta]}$ at server $t$
 is  NULL or not, the user can rewrite it as
 \begin{IEEEeqnarray}{c}\label{expend:answer}
 \mathbf{A}_{t}^{[\theta]}(s)=\mathcal{W}_{\theta}\big((\mathcal{Q}(s,\theta)+t)_n,t\big)+\mathbf{n}_{t}^{[\theta]}(s),~t\in[0:N),s\in[0:k),
 \end{IEEEeqnarray}
where
 \begin{IEEEeqnarray}{c}
 \mathbf{n}_{t}^{[\theta]}(s)\triangleq  \sum\limits_{i\in[0:M)\backslash\{\theta\}}\mathcal{W}_i\big(\mathcal{Q}(s,i),t\big).\notag
 \end{IEEEeqnarray}
This is to say,  if the answer $\mathbf{A}_{t}^{[\theta]}(s)$ is NULL, the user can interpret it as $0$, since in this case,
 $\mathcal{W}_{\theta}\big((\mathcal{Q}(s,\theta)+t)_n,t\big)=0$ and $\mathcal{W}_i\big(\mathcal{Q}(s,i),t\big)=0$ for every $i\in[0:M)\backslash\{\theta\}$.

From \eqref{encoded symbols} and \eqref{supply zero},  $\big(\mathcal{W}_{i}(Q(s,i),0),\ldots,\mathcal{W}_{i}(Q(s,i),N-1)\big)$ constitutes a codeword of the linear $(N,K)$ MDS code, 
so does  $\big(\mathbf{n}_{0}^{[\theta]}(s),\ldots,\mathbf{n}_{N-1}^{[\theta]}(s)\big)$.
 Thus, the reconstruction phase can be depicted as follows.

\vspace{1.5mm}
 \textbf{Reconstruction Phase:} 
Define
 \begin{IEEEeqnarray*}{rCll}
 \Delta_s&\triangleq&\{t\in[0:N)|(\mathcal{Q}(s,\theta)+t)_n\in[n-k:n)\},&~s\in[0:k), \\
 \Lambda_j&\triangleq&\{t\in[0:N)|(\mathcal{Q}(s,\theta)+t)_n=j,s\in[0:k)\},&~j\in[0:n-k).
 \end{IEEEeqnarray*}
\textbf{Step 1.} Given a fixed $s\in[0:k)$, the user recovers the  residual $N-K$ values $\mathbf{n}_{t}^{[\theta]}(s),t\in[0:N)\backslash\Delta_s$, from the $K$ answers $\mathbf{n}_{t}^{[\theta]}(s)=\mathbf{A}_{t}^{[\theta]}(s),t\in\Delta_s$, by the MDS property of the codeword $\big(\mathbf{n}_{0}^{[\theta]}(s),\ldots,\mathbf{n}_{N-1}^{[\theta]}(s)\big)$.

This is because  (i) $\mathcal{W}_{\theta}\big((\mathcal{Q}(s,\theta)+t)_n,t\big)=0$ for $t\in\Delta_s$; (ii)  $|\Delta_s|=d\cdot k=K$ by fact \textbf{F1}, and then the user has all the  desired $K$ values $\mathbf{n}_{t}^{[\theta]}(s)$ from the answers $\mathbf{A}_{t}^{[\theta]}(s)$, $t\in\Delta_s$.

\textbf{Step 2.}  The user cancels the term $\mathbf{n}_{t}^{[\theta]}(s)$ in the answers \eqref{expend:answer} from all severs $t\in[0:N)\backslash\Delta_s$ by those $N-K$ values calculated in Step 1 to obtain
 \begin{equation}\label{decoded contents}
 \{\mathcal{W}_{\theta}\big((\mathcal{Q}(s,\theta)+t)_n,t\big)\,,t\in[0:N)\backslash\Delta_s\},~ s\in[0:k).
 \end{equation}

\textbf{Step 3.} Collecting all the values in \eqref{decoded contents} when $s$ enumerates $[0:k)$, the user respectively reconstructs all the vectors $ \mathcal{W}_{\theta}(j,:)\big(j\in [0:n-k)\big)$ from $\{\mathcal{W}_{\theta}(j,t):t\in\Lambda_j\}$ by means of its MDS property.

This is because given $j\in[0:n-k)$, (i) $\Lambda_{j,s}\triangleq\{t\in[0:N)|(\mathcal{Q}(s,\theta)+t)_n=j\}\subseteq[0:N)\backslash\Delta_s$ is of cardinality $d$ for any fixed $s\in[0:k)$, again by fact  \textbf{F1}; and then (ii) $|\Lambda_j|=d\cdot k=K$ by fact  \textbf{F2}, i.e.,  there are $K$ distinct coded symbols $\{\mathcal{W}_{\theta}(j,t):t\in\Lambda_j\}$ available from \eqref{decoded contents}.

\vspace{1.5mm}
After Reconstruction Phase, the user is able to reconstruct the whole file $ \mathcal{W}_{\theta}$.

\subsection{Proof of Privacy}\label{private lemma}
Denote the set of $k\times M$ matrices with column vectors chosen from $\Omega$ by $\mho$, i.e.,
\begin{IEEEeqnarray}{rCl}
\mho&=&\left\{(\mathbf{q}_0,\mathbf{q}_1,\ldots,\mathbf{q}_{M-1})\;:\;\mathbf{q}_i\in\Omega,\;\forall~i\in[0:M)\right\}. \notag
\end{IEEEeqnarray}
Recall from \eqref{eqn:Q_theta} that $\mathbf{q}_i$ ($i\in[0:M)$) are independent and uniform in $\Omega$. So, according to fact \textbf{F3}, $(\mathbf{q}_{\theta}+t)_n$ is uniform in $\Omega$ for any $\theta\in[0:M)$ and $t\in[0:N)$. Consequently, $\mathbf{q}_0,\ldots,\mathbf{q}_{\theta-1},\left(\mathbf{q}_{\theta}+t\right)_n,\mathbf{q}_{\theta+1},\ldots,\mathbf{q}_{M-1}$ are  independent and uniform in $\Omega$ for any $t\in[0:N)$.
That is, the matrix $\mathcal{Q}_{t}^{[\theta]}$ in \eqref{query formulation} has uniform distribution over $\mho$. Then,
given $\widetilde{\mathcal{Q}}\in\mho$,
\begin{IEEEeqnarray}{rCl}
\Pr(\mathcal{Q}_{t}^{[\theta]}=\widetilde{\mathcal{Q}})=\frac{1}{|\mho|},~\theta\in[0:M),t\in[0:N), \label{Proof of idential distri}
\end{IEEEeqnarray}
which is independent of $\theta$.
Next, for any $t\in[0:N)$ and $\theta\in[0:M)$, we have
\begin{IEEEeqnarray}{rCl}
0&\leq&I(\mathcal{Q}_{t}^{[\theta]},\mathbf{A}_{t}^{[\theta]},\mathbf{y}_{t};\theta) \notag\\
&\leq&I(\mathcal{Q}_{t}^{[\theta]},\mathbf{A}_{t}^{[\theta]},\mathcal{W}_{0:M-1};\theta) \notag\\
&=&I(\mathcal{Q}_{t}^{[\theta]};\theta)+I(\mathcal{W}_{0:M-1};\theta|\mathcal{Q}_{t}^{[\theta]})+I(\mathbf{A}_{t}^{[\theta]};\theta|\mathcal{Q}_{t}^{[\theta]},\mathcal{W}_{0:M-1}) \notag\\
&\overset{(a)}{=}&H(\mathcal{W}_{0:M-1}|\mathcal{Q}_{t}^{[\theta]})-H(\mathcal{W}_{0:M-1}|\theta,\mathcal{Q}_{t}^{[\theta]}) \notag \\
&\overset{(b)}{=}&H(\mathcal{W}_{0:M-1})-H(\mathcal{W}_{0:M-1}) \notag\\
&=&0, \notag
\end{IEEEeqnarray}
where $(a)$ is because the query  is independent of $\theta$ by \eqref{Proof of idential distri} and the answer is a determined function of the received query and the files by \eqref{server answer} such that  $I(\mathcal{Q}_{t}^{[\theta]};\theta)=0$ and  $I(\mathbf{A}_{t}^{[\theta]};\theta|\mathcal{Q}_{t}^{[\theta]},\mathcal{W}_{0:M-1})=0$; $(b)$ is due to the fact that the files are independent of  the desired file index and the query.

Thus, privacy  of  the new PIR scheme follows from \eqref{infor-theoretic privacy}.

\subsection{Proof of Performance}\label{performance lemma}

Recall that $\ell_t$ is  the  answer length of the query $\mathcal{Q}_{t}^{[\theta]}$ at server $t$, clearly which is
\begin{eqnarray*}
\ell_t=\sum_{s=0}^{k-1}\ell( \mathbf{A}_{t}^{[\theta]}(s)),
\end{eqnarray*}
where $\ell(\mathbf{A}_{t}^{[\theta]}(s))$ is the length of $\mathbf{A}_{t}^{[\theta]}(s)$, satisfying
\begin{equation}\label{Ats:length}
\ell(\mathbf{A}_{t}^{[\theta]}(s))=\begin{cases}
                                0, & \text{if}~\mathcal{Q}_{t}^{[\theta]}(s,:)\in[n-k:n)^M \\
                                1, & \text{otherwise}
                             \end{cases},~s\in[0:k)
\end{equation}
by \eqref{server answer}.

It follows from fact  \textbf{F3} and \eqref{Ats:length} that
\begin{IEEEeqnarray}{c}
\mathbb{E}\left[\ell(\mathbf{A}_{t}^{[\theta]}(s))\right]=1\cdot \Pr(\ell(\mathbf{A}_{t}^{[\theta]}(s))=1)+0\cdot \Pr(\ell(\mathbf{A}_{t}^{[\theta]}(s))=0)=1-\left(\frac{k}{n}\right)^M. \notag
\end{IEEEeqnarray}
Then, we have
\begin{IEEEeqnarray}{rCl}\label{Eqn_D}
D&=&\mathbb{E}\left[\sum_{t=0}^{N-1}\ell_t\right]\notag\\
&=&\mathbb{E}\left[\sum_{t=0}^{N-1}\sum_{s=0}^{k-1}\ell(\mathbf{A}_{t}^{[\theta]}(s))\right]\notag\\
&=&\sum_{t=0}^{N-1}\sum_{s=0}^{k-1}\mathbb{E}\left[\ell(\mathbf{A}_{t}^{[\theta]}(s))\right]\notag\\
&=&Nk\left(1-\left(\frac{k}{n}\right)^M\right).
\end{IEEEeqnarray}

Finally, substituting  \eqref{subpacketization}, \eqref{Eqn_D} and the fact $\frac{K}{N}=\frac{k}{n}$ into \eqref{Moded rate}, we obtain
\begin{IEEEeqnarray}{rCl}
R&=&\frac{L}{D}\notag\\
&=&\frac{K(n-k)}{Nk\left(1-\left(\frac{k}{n}\right)^M\right)}\notag\\
&=&\left(1+\frac{K}{N}+\left(\frac{K}{N}\right)^2+\ldots+\left(\frac{K}{N}\right)^{M-1}\right)^{-1}, \notag
\end{IEEEeqnarray}
which achieves the capacity of the coded PIR scheme in \eqref{the capacity}.

\section{The Optimality on The File Length}\label{proof of lower bound}\label{the converse bound}
\subsection{Some Necessary Conditions for Capacity-achieving Coded linear PIR Schemes}
The capacity of coded PIR has been determined in \cite{Ulukus}. In this subsection, we provide a detailed analysis of the converse proof to emphasize on three necessary conditions  \textbf{P1}-\textbf{P3} for capacity-achieving coded linear PIR schemes.
The similar approach was used in \cite{Tian and Sun} for the replicated PIR scenario. The proofs of Lemma \ref{answer independence}-\ref{intial step} are left in the Appendix.

\begin{Lemma}\label{answer independence}
For any $\Gamma\subseteq[0:N),|\Gamma|=K,\Lambda\subseteq[0:M),\theta\in[0:M)$, the answers from servers in $\Gamma$ are mutually and statistically independent conditioned on the files $\mathcal{W}_{\Lambda}$, i.e.,
\begin{equation}\label{Lem:answer indenp}
H(\mathbf{A}_{\Gamma}^{[\theta]}|\mathcal{W}_{\Lambda},\mathcal{Q}_{\Gamma}^{[\theta]})=\sum\limits_{t\in\Gamma}H(\mathbf{A}_{t}^{[\theta]}|\mathcal{W}_{\Lambda},\mathcal{Q}_{t}^{[\theta]}),~\theta\in[0:M).
\end{equation}
Equivalently, an $(N,K,M)$ coded  linear PIR scheme must have
\begin{enumerate}
  \item[\textbf{P1: }] The answers from servers in $\Gamma$ are statistically independent conditioned on the files $\mathcal{W}_{\Lambda}$ for any query realization $\widetilde{\mathcal{Q}}_{0:N-1}^{[\theta]}$  with positive probability, i.e.,
                        \begin{equation}\label{P1}
                        H(\mathbf{A}_{\Gamma}^{[\theta]}|\mathcal{W}_{\Lambda},\mathcal{Q}_{\Gamma}^{[\theta]}=\widetilde{\mathcal{Q}}_{\Gamma}^{[\theta]})=\sum\limits_{t\in\Gamma}H(\mathbf{A}_{t}^{[\theta]}|\mathcal{W}_{\Lambda},\mathcal{Q}_{t}^{[\theta]}=\widetilde{\mathcal{Q}}_{t}^{[\theta]}),~\theta\in[0:M).
                        \end{equation}
\end{enumerate}
\end{Lemma}
\begin{Lemma}\label{recursion lemma}
For any $\Lambda\subseteq[0:M),\theta\in\Lambda,\theta'\in[0:M)\backslash\Lambda,\Gamma\subseteq[0:N),|\Gamma|=K$,
\begin{equation}\label{Lemma recursion}
H(\mathbf{A}_{0:N-1}^{[\theta]}|\mathcal{W}_{\Lambda},\mathcal{Q}_{0:N-1}^{[\theta]})\geq\frac{KL}{N}+\frac{K}{N}H(\mathbf{A}_{0:N-1}^{[\theta']}|\mathcal{W}_{\Lambda},\mathcal{W}_{\theta'},\mathcal{Q}_{0:N-1}^{[\theta']}).
\end{equation}
Moreover,  to establish the equality in \eqref{Lemma recursion}, the $(N,K,M)$ coded  linear PIR scheme must have
\begin{enumerate}
  \item[\textbf{P2: }] The answers in any $K$ servers determine all the answers conditioned on the files $\mathcal{W}_{\Lambda}$ for  any query realization $\widetilde{\mathcal{Q}}_{0:N-1}^{[\theta]}$  with positive probability, i.e.,
                        \begin{equation}\label{P2}
                        H(\mathbf{A}_{0:N-1}^{[\theta]}|\mathcal{W}_{\Lambda},\mathcal{Q}_{0:N-1}^{[\theta]}=\widetilde{\mathcal{Q}}_{0:N-1}^{[\theta]})= H(\mathbf{A}_{\Gamma}^{[\theta]}|\mathcal{W}_{\Lambda},\mathcal{Q}_{\Gamma}^{[\theta]}=\widetilde{\mathcal{Q}}_{\Gamma}^{[\theta]}),\quad\theta\in\Lambda.
                        \end{equation}
\end{enumerate}
\end{Lemma}

\begin{Lemma}\label{intial step}
For any  $(N,K,M)$ coded  linear PIR scheme,
\begin{equation}\label{Lemma intial}
H(\mathbf{A}_{0:N-1}^{[\theta]}|\mathcal{W}_{\theta},\mathcal{Q}_{0:N-1}^{[\theta]})\leq\frac{L}{R}-L.
\end{equation}
Moreover, to establish the equality in \eqref{Lemma intial}, the $(N,K,M)$ coded linear PIR scheme must have
\begin{enumerate}
  \item[\textbf{P3: }] The $N$ answers are mutually independent for any query realization $\widetilde{\mathcal{Q}}_{0:N-1}^{[\theta]}$ with positive probability, i.e.,
                        \begin{equation}\label{download cost}
                        D_{rel}=H(\mathbf{A}_{0:N-1}^{[\theta]}|\mathcal{Q}_{0:N-1}^{[\theta]}=\widetilde{\mathcal{Q}}_{0:N-1}^{[\theta]})= \sum\limits_{t=0}^{N-1}H(\mathbf{A}_{t}^{[\theta]}|\mathcal{Q}_{t}^{[\theta]}=\widetilde{\mathcal{Q}}_{t}^{[\theta]}),\quad\theta\in[0:M),
                        \end{equation}
                        where $D_{rel}$  denotes the download cost  for the query realization $\widetilde{\mathcal{Q}}_{0:N-1}^{[\theta]}$.
\end{enumerate}
\end{Lemma}

\begin{Theorem}
Any capacity-achieving coded linear PIR scheme must satisfy the three necessary conditions \emph{\textbf{P1}}-\emph{\textbf{P3}}.
\end{Theorem}

\begin{IEEEproof}  First,  \textbf{P1} is satisfied for any  coded linear PIR scheme by Lemma \ref{answer independence}.

Next, set $[0:M)\backslash\theta=\{\theta_1,\ldots, \theta_{M-1}\}$.
Then, by \eqref{Lemma intial},
\begin{IEEEeqnarray}{rCl}
\frac{L}{R}-L&\geq&H(\mathbf{A}_{0:N-1}^{[\theta]}|\mathcal{W}_{\theta},\mathcal{Q}_{0:N-1}^{[\theta]})\notag\\
&\overset{(a)}{\geq}&L(\frac{K}{N}+\frac{K^2}{N^2}+\ldots+\frac{K^{M-1}}{N^{M-1}})+\frac{K^{M-1}}{N^{M-1}}H(\mathbf{A}_{0:N-1}^{[\theta_{M-1}]}|\mathcal{W}_{0:M-1},\mathcal{Q}_{0:N-1}^{[\theta_{M-1}]})\notag\\
&\overset{(b)}{=}&L(\frac{K}{N}+\frac{K^2}{N^2}+\ldots+\frac{K^{M-1}}{N^{M-1}}),\label{upper bound}
\end{IEEEeqnarray}
where $(a)$ follows from recursively applying Lemma \ref{recursion lemma}  to $\theta'=\theta_{i}$, $i=1,\ldots, M-1$; $(b)$ is because of \eqref{answer function}, i.e.,
$H(\mathbf{A}_{0:N-1}^{[\theta_{M-1}]}|\mathcal{W}_{0:M-1},$ $\mathcal{Q}_{0:N-1}^{[\theta_{M-1}]})=0$.

According to the capacity $C$ in \eqref{the capacity}, a capacity-achieving $(N,K,M)$ coded linear PIR scheme achieves the quality  in \eqref{upper bound}. Consequently,  any capacity-achieving $(N,K,M)$ coded linear PIR scheme must attain the equalities
 \eqref{Lemma recursion} and \eqref{Lemma intial}, i.e., \textbf{P2} and \textbf{P3} are necessary conditions for such PIR scheme.
\end{IEEEproof}

\subsection{Rank Properties of the Answer-Interference Matrix of Capacity-Achieving Coded Linear PIR Schemes}

\begin{Lemma} (\cite[Lemma 8]{Zhang2})\label{entroy and rank}
In an $(N,K,M)$ coded linear PIR scheme, for any $\Lambda\subseteq[0:M),\theta\in[0:M)$, and any realization $\widetilde{\mathcal{Q}}_{0:N-1}^{[\theta]}$ of the random queries $\mathcal{Q}_{0:N-1}^{[\theta]}$, we have
\begin{equation}
H(\mathbf{A}_{t}^{[\theta]}|\mathcal{W}_{\Lambda},\mathcal{Q}_{t}^{[\theta]}=\widetilde{\mathcal{Q}}_{t}^{[\theta]})=\mathrm{rank}(\widetilde{\mathcal{A}}_{t,[0:M)\backslash\Lambda}^{[\theta]}),\quad t\in[0:N), \notag
\end{equation}
where $\widetilde{\mathcal{A}}_{t,[0:M)}^{[\theta]}=\big( \widetilde{\mathcal{A}}_{t,0}^{[\theta]},\widetilde{\mathcal{A}}_{t,1}^{[\theta]},\ldots,\widetilde{\mathcal{A}}_{t,M-1}^{[\theta]}\big)$  is the answer matrix of query realization $\widetilde{\mathcal{Q}}_{t}^{[\theta]}$ at server $t$.
\end{Lemma}


In  the following, Lemma \ref{D and L relation}  determines the relation between the \textit{specific} download cost  and the file length for any concrete  realization $\widetilde{\mathcal{Q}}_{0:N-1}^{[\theta]}$ of the random queries $\mathcal{Q}_{0:N-1}^{[\theta]}$.
It is worth pointing out  that  the same relation between the \textit{average} download cost and the file length for random queries $\mathcal{Q}_{0:N-1}^{[\theta]}$ has been established in \cite{Zhang}.
Essentially,  Lemma \ref{D and L relation}  implies the result in \cite{Zhang} by the fact that the relation holds for all  the specific download cost and so does for average ones,
but not vice versa.

\begin{Lemma}\label{D and L relation}
In a capacity-achieving $(N,K,M)$ coded linear PIR scheme, for any realization $\widetilde{\mathcal{Q}}_{0:N-1}^{[\theta]}$ of the random queries $\mathcal{Q}_{0:N-1}^{[\theta]}$ with positive probability, $\theta\in[0:M)$, all the answer-interference matrices $\widetilde{\mathcal{A}}_{t,[0:M)\backslash\{\theta\}}^{[\theta]}(t\in[0:N))$ have the same rank $r$, i.e.,
\begin{IEEEeqnarray}{c}
r=\mathrm{rank}(\widetilde{\mathcal{A}}_{0,[0:M)\backslash\{\theta\}}^{[\theta]})=\ldots=\mathrm{rank}(\widetilde{\mathcal{A}}_{N-1,[0:M)\backslash\{\theta\}}^{[\theta]}),\label{D-L3}
\end{IEEEeqnarray}
and the download cost satisfies
\begin{equation}\label{Eqn_Dr}
D_{rel}-L=K\cdot r.
\end{equation}
\end{Lemma}
\begin{IEEEproof}
For any realization $\widetilde{\mathcal{Q}}_{0:N-1}^{[\theta]}$ of the random queries $\mathcal{Q}_{0:N-1}^{[\theta]}$, we have
\begin{IEEEeqnarray}{rCl}
L&=&H(\mathcal{W}_{\theta})\notag\\
&\overset{(a)}{=}&H(\mathcal{W}_{\theta}|\mathcal{Q}_{0:N-1}^{[\theta]}=\widetilde{\mathcal{Q}}_{0:N-1}^{[\theta]})-H(\mathcal{W}_{\theta}|\mathbf{A}_{0:N-1}^{[\theta]},\mathcal{Q}_{0:N-1}^{[\theta]}=\widetilde{\mathcal{Q}}_{0:N-1}^{[\theta]})\notag\\
&=&I(\mathcal{W}_{\theta};\mathbf{A}_{0:N-1}^{[\theta]}|\mathcal{Q}_{0:N-1}^{[\theta]}=\widetilde{\mathcal{Q}}_{0:N-1}^{[\theta]})\notag\\
&=&H(\mathbf{A}_{0:N-1}^{[\theta]}|\mathcal{Q}_{0:N-1}^{[\theta]}=\widetilde{\mathcal{Q}}_{0:N-1}^{[\theta]})-H(\mathbf{A}_{0:N-1}^{[\theta]}|\mathcal{W}_{\theta},\mathcal{Q}_{0:N-1}^{[\theta]}=\widetilde{\mathcal{Q}}_{0:N-1}^{[\theta]})\notag\\
&\overset{(b)}{=}&D_{rel}-H(\mathbf{A}_{0:N-1}^{[\theta]}|\mathcal{W}_{\theta},\mathcal{Q}_{0:N-1}^{[\theta]}=\widetilde{\mathcal{Q}}_{0:N-1}^{[\theta]}),\label{download L}
\end{IEEEeqnarray}
where $(a)$ follows from \eqref{query independence} and \eqref{correctness constrain}, i.e.,
$H(\mathcal{W}_{\theta})=H(\mathcal{W}_{\theta}|\mathcal{Q}_{0:N-1}^{[\theta]}=\widetilde{\mathcal{Q}}_{0:N-1}^{[\theta]})$ and
$H(\mathcal{W}_{\theta}|\mathbf{A}_{0:N-1}^{[\theta]},\mathcal{Q}_{0:N-1}^{[\theta]}=\widetilde{\mathcal{Q}}_{0:N-1}^{[\theta]})=0$; $(b)$ is due to \eqref{download cost}.

Given any $\Gamma\subseteq[0:N),|\Gamma|=K,i\in\Gamma$, by \eqref{P1} and \eqref{P2}, we have
\begin{IEEEeqnarray}{rCl}
H(\mathbf{A}_{0:N-1}^{[\theta]}|\mathcal{W}_{\theta},\mathcal{Q}_{0:N-1}^{[\theta]}=\widetilde{\mathcal{Q}}_{0:N-1}^{[\theta]})
&=&H(\mathbf{A}_{\Gamma}^{[\theta]}|\mathcal{W}_{\theta},\mathcal{Q}_{\Gamma}^{[\theta]}=\widetilde{\mathcal{Q}}_{\Gamma}^{[\theta]}) \notag\\
&=&\sum\limits_{t\in\Gamma}H(\mathbf{A}_{t}^{[\theta]}|\mathcal{W}_{\theta},\mathcal{Q}_{t}^{[\theta]}=\widetilde{\mathcal{Q}}_{t}^{[\theta]}).\label{D-L1}
\end{IEEEeqnarray}
Further replacing $\Gamma$ in \eqref{D-L1} by  $\Gamma'=(\Gamma\backslash\{i\})\cup\{j\}$ for  any $j\in[0:N)\backslash\Gamma$,  we have
\begin{IEEEeqnarray}{rCl}
H(\mathbf{A}_{0:N-1}^{[\theta]}|\mathcal{W}_{\theta},\mathcal{Q}_{0:N-1}^{[\theta]}=\widetilde{\mathcal{Q}}_{0:N-1}^{[\theta]})&=&\sum\limits_{t\in\Gamma'}H(\mathbf{A}_{t}^{[\theta]}|\mathcal{W}_{\theta},\mathcal{Q}_{t}^{[\theta]}=\widetilde{\mathcal{Q}}_{t}^{[\theta]}). \label{D-L2}
\end{IEEEeqnarray}
Then, combining \eqref{D-L1} and \eqref{D-L2}, we obtain $\sum\limits_{t\in\Gamma}H(\mathbf{A}_{t}^{[\theta]}|\mathcal{W}_{\theta},\mathcal{Q}_{t}^{[\theta]}=\widetilde{\mathcal{Q}}_{t}^{[\theta]})=\sum\limits_{t\in\Gamma'}H(\mathbf{A}_{t}^{[\theta]}|\mathcal{W}_{\theta},\mathcal{Q}_{t}^{[\theta]}=\widetilde{\mathcal{Q}}_{t}^{[\theta]})$. The definitions of the two sets $\Gamma$ and $\Gamma'$  lead to
\begin{equation}\label{interference entropy}
H(\mathbf{A}_{i}^{[\theta]}|\mathcal{W}_{\theta},\mathcal{Q}_{i}^{[\theta]}=\widetilde{\mathcal{Q}}_{i}^{[\theta]})=H(\mathbf{A}_{j}^{[\theta]}|\mathcal{W}_{\theta},\mathcal{Q}_{j}^{[\theta]}=\widetilde{\mathcal{Q}}_{j}^{[\theta]})
\end{equation}
for any $i,j\in[0:N),i\neq j$.

Therefore, applying Lemma \ref{entroy and rank}  to \eqref{interference entropy}, we reach \eqref{D-L3}, which gives
$H(\mathbf{A}_{0:N-1}^{[\theta]}|\mathcal{W}_{\theta},\mathcal{Q}_{0:N-1}^{[\theta]}=\widetilde{\mathcal{Q}}_{0:N-1}^{[\theta]})=K\cdot r$ by  \eqref{D-L1}
and then \eqref{Eqn_Dr} by \eqref{download L}.
\end{IEEEproof}

\subsection{A Lower Bound on the Minimum File Length of Capacity-Achieving Coded Linear PIR Schemes}

In this subsection, we establish an information theoretic lower bound on the minimum file length among all capacity-achieving $(N,K,M)$ coded linear PIR schemes.

 \begin{Lemma}\label{lem_LNK}
For a capacity-achieving $(N,K,M)$ coded linear PIR scheme and an integer $\theta\in[0:M)$, let $\widetilde{\mathcal{Q}}_{0:N-1}^{[\theta]}$ be a query
realization  with positive probability, and denote  $r$  the rank of all its answer-interference matrices $\widetilde{\mathcal{A}}_{t,[0:M)\backslash\{\theta\}}^{[\theta]}$, $t\in[0:N)$.
Then, its file length  satisfies
\begin{eqnarray}
L\ge (N-K)\cdot r. \notag
\end{eqnarray}
\end{Lemma}

\begin{IEEEproof}
For the realization $\widetilde{\mathcal{Q}}_{0:N-1}^{[\theta]}$, we have
\begin{IEEEeqnarray}{rCl}\label{D>Nr2}
K\cdot r+L&\overset{(a)}{=}&D_{rel}\nonumber\\
&\overset{(b)}{=}&\sum\limits_{t=0}^{N-1}H(\mathbf{A}_{t}^{[\theta]}|\mathcal{Q}_{t}^{[\theta]}=\widetilde{\mathcal{Q}}_{t}^{[\theta]})\nonumber\\
&\overset{(c)}{=}&\sum\limits_{t=0}^{N-1}\mathrm{rank}(\widetilde{\mathcal{A}}_{t,[0:M)}^{[\theta]})\nonumber\\
&\overset{(d)}{\geq}&\sum\limits_{t=0}^{N-1}\mathrm{rank}(\widetilde{\mathcal{A}}_{t,[0:M)\backslash\{\theta\}}^{[\theta]})\nonumber\\
&=&N\cdot r,
\end{IEEEeqnarray}
where $(a)$ and $(b)$ respectively  follow from  \eqref{Eqn_Dr} and \eqref{download cost}; $(c)$ is due to Lemma \ref{entroy and rank}; $(d)$ is because the rank of matrix is not less than the rank of its sub-matrix, i.e.,
 $\mathrm{rank}(\widetilde{\mathcal{A}}_{t,[0:M)}^{[\theta]})\geq\mathrm{rank}(\widetilde{\mathcal{A}}_{t,[0:M)\backslash\{\theta\}}^{[\theta]})=r$. Therefore, we get $L\ge (N-K)\cdot r$.
\end{IEEEproof}

Now, we are ready to derive our bound based on  the following observation from  the privacy constraint in \eqref{idential distributed} and  the definition of coded linear PIR scheme in \eqref{linear answer function}.

\begin{adjustwidth}{20pt}{0cm}
\textbf{Observation:} For any  query realization $\widetilde{\mathcal{Q}}_{0:N-1}^{[\theta]}$ with positive probability,
the query $\widetilde{\mathcal{Q}}_{t}^{[\theta]}$, which is sent by the user to the  server $t$ for retrieving file $\mathcal{W}_{\theta}$, can also be sent to the same server but for retrieving
a distinct file $\mathcal{W}_{\theta'}$ in another query realization $\widetilde{\mathcal{Q}}_{0:N-1}^{[\theta']}$ with positive probability, i.e., $\widetilde{\mathcal{Q}}_{t}^{[\theta]}=\widetilde{\mathcal{Q}}_{t}^{[\theta']}$.  As a result, in the two
query realizations $\widetilde{\mathcal{Q}}_{0:N-1}^{[\theta]}$ and $\widetilde{\mathcal{Q}}_{0:N-1}^{[\theta']}$, sever $t$ will respond with
the same answer matrix  $\widetilde{\mathcal{A}}_{t,[0:M)}^{[\theta']}=\widetilde{\mathcal{A}}_{t,[0:M)}^{[\theta]}$.
\end{adjustwidth}

\vspace{1.5mm}
\begin{adjustwidth}{20pt}{0cm}
The former is because the two  queries $\mathcal{Q}_{t}^{[\theta]}$ and $\mathcal{Q}_{t}^{[\theta']}$ have the  identical distribution for any two distinct desired $\theta,\theta'\in [0:M)$.
Whereas, the latter is because
 the matrix $\mathcal{A}_{t,[0:M)}^{[\theta]}$ ($\mathcal{A}_{t,[0:M)}^{[\theta']}$) is completely determined to be $\widetilde{\mathcal{A}}_{t,[0:M)}^{[\theta]}$ ($\widetilde{\mathcal{A}}_{t,[0:M)}^{[\theta']}$) by  the received  query  realization $\widetilde{\mathcal{Q}}_{t}^{[\theta]}$ ($\widetilde{\mathcal{Q}}_{t}^{[\theta']}$).
\end{adjustwidth}

\begin{Theorem}\label{The_general bound}
Given any $N,K$ and $M$ with $N>K$ and $M>1$, the file length of any capacity-achieving $(N,K,M)$ coded linear PIR scheme satisfies
\begin{eqnarray}\label{Eqn_LNK}
L\geq N-K.
\end{eqnarray}
\end{Theorem}
\begin{IEEEproof}
Assume that \eqref{Eqn_LNK} is not true for a capacity-achieving $(N,K,M)$ coded linear PIR scheme. Then, it follows from Lemmas \ref{D and L relation} and \ref{lem_LNK} that
all the answer-interference matrices  will have the rank $r= 0$ for  each of its query
realization with positive probability.

Let $\mathcal{W}_{\theta}$ ($\theta\in[0:M)$) be a desired file  and $\widetilde{\mathcal{Q}}_{0:N-1}^{[\theta]}$ be  a query realization with positive probability. In this scenario, to ensure that the user can decode the desired file $\mathcal{W}_{\theta}$, the user must directly download the  file  from the servers, i.e., there exists at least one server $t\in[0:N)$ such that its answer matrix
\begin{IEEEeqnarray}{c}
\widetilde{\mathcal{A}}_{t,[0:M)}^{[\theta]}=\left( \widetilde{\mathcal{A}}_{t,0}^{[\theta]},\ldots,\widetilde{\mathcal{A}}_{t,\theta}^{[\theta]},\ldots,\widetilde{\mathcal{A}}_{t,M-1}^{[\theta]} \right) \notag
\end{IEEEeqnarray}
satisfies that  $\widetilde{\mathcal{A}}_{t,[0:M)\backslash\{\theta\}}^{[\theta]}=\mathbf{0}$  and $\widetilde{\mathcal{A}}_{t,\theta}^{[\theta]}\neq\mathbf{0}$.

By the \textbf{Observation}, for any server $t$ and any $\theta'\neq\theta$, there exists a  query realization $\widetilde{Q}_t^{[\theta']}$ with positive probability such that  the sever $t$ will respond with
the same answer matrix, i.e., $\widetilde{\mathcal{A}}_{t,[0:M)}^{[\theta']}=\widetilde{\mathcal{A}}_{t,[0:M)}^{[\theta]}$. Then, the answer-interference matrix
\begin{IEEEeqnarray}{c}
\widetilde{\mathcal{A}}_{t,[0:M)\backslash\{\theta'\}}^{[\theta']}=\widetilde{\mathcal{A}}_{t,[0:M)\backslash\{\theta'\}}^{[\theta]}=\left( \widetilde{\mathcal{A}}_{t,0}^{[\theta]},\ldots,\widetilde{\mathcal{A}}_{t,\theta'-1}^{[\theta]},\widetilde{\mathcal{A}}_{t,\theta'+1}^{[\theta]},\ldots,\widetilde{\mathcal{A}}_{t,M-1}^{[\theta]} \right)\notag
\end{IEEEeqnarray}
with
\begin{IEEEeqnarray}{c}
r=\mathrm{rank}(\widetilde{\mathcal{A}}_{t,[0:M)\backslash\{\theta'\}}^{[\theta']})\geq 1, \notag
\end{IEEEeqnarray}
which contradicts the assumption that  the rank  $r = 0$ for any query realization with
positive probability. This completes the proof.
\end{IEEEproof}

\begin{Remark}
When $K=1$, the $(N,1,M)$ coded PIR scheme is just a repetition scheme. In this case, our bound becomes $L\ge N-1$, which is
consistent with the bound in \cite{Tian and Sun}.
\end{Remark}

Specifically, when $M>\lfloor \frac{K}{\gcd(N,K)}-\frac{K}{N-K}\rfloor+1$, we can further improve the lower bound to a tight one.

\begin{Theorem}\label{The_tight bound}
Given any $N,K$ and $M$ with $N>K$ and $M>\big\lfloor \frac{K}{\gcd(N,K)}-\frac{K}{N-K}\big\rfloor+1$, the file length of any capacity-achieving $(N,K,M)$ coded linear PIR scheme satisfies
\begin{eqnarray*}
L\geq \frac{K(N-K)}{\gcd(N,K)}.
\end{eqnarray*}
\end{Theorem}

\begin{IEEEproof}
Recall that $n=\frac{N}{\gcd(N,K)},k=\frac{K}{\gcd(N,K)}$. Suppose that $L=K\cdot \lambda < \frac{K(N-K)}{\gcd(N,K)}$, i.e., $\lambda <n-k$. Let $\widetilde{\mathcal{Q}}_{0:N-1}^{[\theta]}$ ($\theta\in[0:M)$) be a realization of the random queries $\mathcal{Q}_{0:N-1}^{[\theta]}$ with positive probability.
Then,  all the answer-interference matrices have the same rank $r$ by Lemma \ref{D and L relation}. Using $L=K\cdot\lambda$ in \eqref{D>Nr2}, we have
\begin{equation}\label{r--r+1}
 r\overset{(e)}{\leq}\frac{K\cdot\lambda }{N-K}=\frac{k\cdot\lambda }{n-k},
\end{equation}
which results in
\begin{equation}\label{r<k}
r\leq \left\lfloor\frac{k\cdot(n-k-1)}{n-k}\right\rfloor=\left\lfloor k-\frac{k}{n-k}\right\rfloor
\end{equation}
by the fact that the rank $r$ must be an integer. In addition, note that, for any $\lambda <n-k$, $\frac{k\cdot\lambda}{n-k}$ cannot be an integer since $\gcd(k,n-k)=1$.
This is to say, the inequality in  $(e)$ of \eqref{r--r+1} always holds, and so does for the one in $(d)$ of \eqref{D>Nr2}, i.e.,
\begin{equation}\label{D neq Nr}
\sum\limits_{t=0}^{N-1}\mathrm{rank}(\widetilde{\mathcal{A}}_{t,[0:M)}^{[\theta]})>N\cdot r.
\end{equation}
Again by $\mathrm{rank}(\widetilde{\mathcal{A}}_{t,[0:M)}^{[\theta]})\geq\mathrm{rank}(\widetilde{\mathcal{A}}_{t,[0:M)\backslash\{\theta\}}^{[\theta]})=r$,  \eqref{D neq Nr}
indicates that there exists at least one server $t'\in[0:N)$ whose answer matrix satisfies
\begin{equation}
\mathrm{rank}(\widetilde{\mathcal{A}}_{t',[0:M)}^{[\theta]})\geq r+1. \notag
\end{equation}

Next, we write the matrix $\widetilde{\mathcal{A}}_{t',[0:M)\backslash\{\theta\}}^{[\theta]}$ as $M-1$ sub-matrix, i.e.,
\begin{eqnarray*}
\widetilde{\mathcal{A}}_{t',[0:M)\backslash\{\theta\}}^{[\theta]}&=&\left( \widetilde{\mathcal{A}}_{t',0}^{[\theta]},\ldots,\widetilde{\mathcal{A}}_{t',\theta-1}^{[\theta]},\widetilde{\mathcal{A}}_{t',\theta+1}^{[\theta]},\ldots,\widetilde{\mathcal{A}}_{t',M-1}^{[\theta]} \right),
\end{eqnarray*}
which has the rank $r<M-1$ by \eqref{r<k} if  $M>\big\lfloor k-\frac{k}{n-k}\big\rfloor+1$. Therefore,  there must be a $\theta'\in[0:M)\backslash\{\theta\}$ such that every column in $\widetilde{\mathcal{A}}_{t',\theta'}^{[\theta]}$
is a linear combination of the ones in $\widetilde{\mathcal{A}}_{t',[0:M)\backslash\{\theta,\theta'\}}^{[\theta]}$, which implies
\begin{equation}\label{recursion relation}
\mathrm{rank}(\widetilde{\mathcal{A}}_{t',[0:M)\backslash\{\theta'\}}^{[\theta]})=
\mathrm{rank}(\widetilde{\mathcal{A}}_{t',[0:M)}^{[\theta]})\geq r+1.
\end{equation}

By the \textbf{Observation}, there exists a new query realization $\widetilde{\mathcal{Q}}_{0:N-1}^{[\theta']}$ with positive probability such that the $t'$-th sever  will respond with
the same answer matrices, i.e., $\widetilde{\mathcal{A}}_{t',[0:M)}^{[\theta']}=\widetilde{\mathcal{A}}_{t',[0:M)}^{[\theta]}$.
Then, the answer-interference matrix $\widetilde{\mathcal{A}}_{t',[0:M)\backslash\{\theta'\}}^{[\theta']}$ satisfies
\begin{equation}
\mathrm{rank}(\widetilde{\mathcal{A}}_{t',[0:M)\backslash\{\theta'\}}^{[\theta']})=\mathrm{rank}(\widetilde{\mathcal{A}}_{t',[0:M)\backslash\{\theta'\}}^{[\theta]})\geq r+1 \notag
\end{equation}
by \eqref{recursion relation}.
Again by  Lemma \ref{D and L relation}, all the answer-interference matrices $\widetilde{\mathcal{A}}_{t',[0:M)\backslash\{\theta'\}}^{[\theta']}$ ($t'\in [0:N)$)
have the same rank $r+1$. That is,  the query realization $\widetilde{\mathcal{Q}}_{0:N-1}^{[\theta']}$ still satisfies all the prerequisites. Hence, we can  repeat the above procedures
till
\begin{equation*}
\mathrm{rank}(\widetilde{\mathcal{A}}_{t^*,[0:M)\backslash\{\theta^*\}}^{[\theta^*]})>\left\lfloor k-\frac{k}{n-k}\right\rfloor
\end{equation*}
for some sever index $t^*\in [0:N)$ and file index $\theta^*\in [0:M)$, which contradicts \eqref{r<k}. Therefore, we arrive at the conclusion.
\end{IEEEproof}

Obviously,  Theorem  \ref{The_tight bound} degrades to the result in Theorem  \ref{The_general bound} in terms of the lower bound on the minimum file length if $K|N$.
\begin{Corollary}
 Given any $N,K$ and $M$ with $M>1$, the file length of the new coded linear PIR scheme achieves the lower bound $N-K$ on the file length of capacity-achieving coded linear PIR schemes for $K|N$.
\end{Corollary}

\section{Conclusion}\label{conclusion}
 In this paper, for the setting with MDS coded servers, we considered the problem of minimizing file length among all capacity-achieving coded linear PIR schemes. Firstly, we proposed a new capacity-achieving $(N,K,M)$ coded scheme with the file length  $\frac{K(N-K)}{\gcd(N,K)}$, which has dramatically reduced the file length required for capacity-achieving PIR schemes in the literature. Secondly, we derived lower bound on the minimum file length for capacity-achieving coded linear PIR schemes. With respect to the  bound,
 the file length of our scheme is shown to be optimal if $M>\big\lfloor \frac{K}{\gcd(N,K)}-\frac{K}{N-K}\big\rfloor+1$, and  be within a multiplicative gap $\frac{K}{\gcd(N,K)}$ of the lower bound in the other case.

 In this sense,  the problem of minimizing file length remains open for  $M\le \big\lfloor \frac{K}{\gcd(N,K)}-\frac{K}{N-K}\big\rfloor+1$,
 which deserves further studies in the future.

\section*{Appendix}
Before presenting the proofs of Lemma \ref{answer independence}-\ref{intial step}, we prove two lemmas. Though they  can be proved similarly to  \cite[Lemma 2]{Zhang2}, we briefly summarize their proofs for the completeness.

\begin{Lemma}\label{appendix privacy}
For any $\Lambda\subseteq[0:M)$,
\begin{IEEEeqnarray}{c}
H(\mathbf{A}_{t}^{[\theta]}|\mathcal{W}_{\Lambda},\mathcal{Q}_{t}^{[\theta]})=H(\mathbf{A}_{t}^{[\theta']}|\mathcal{W}_{\Lambda},\mathcal{Q}_{t}^{[\theta']}),~\forall \theta,\theta'\in[0:M),t\in[0:N).
\label{Lemma:privacy}
\end{IEEEeqnarray}
\end{Lemma}
\begin{IEEEproof}
From the privacy constraint in \eqref{infor-theoretic privacy}, for any $\theta\in[0:M)$, we have
\begin{IEEEeqnarray}{rCl}
0&=&I(\mathcal{Q}_t^{[\theta]},\mathbf{A}_{t}^{[\theta]},\mathbf{y}_t;\theta) \notag \\
&\geq&I(\mathcal{Q}_t^{[\theta]};\theta) \notag \\
&\overset{(a)}{=}&I(\mathcal{Q}_{t}^{[\theta]};\theta)+I(\mathcal{W}_{0:M-1};\theta|\mathcal{Q}_{t}^{[\theta]})+I(\mathbf{A}_{t}^{[\theta]};\theta|\mathcal{Q}_{t}^{[\theta]},\mathcal{W}_{0:M-1}) \notag\\
&=&I(\mathcal{Q}_{t}^{[\theta]},\mathbf{A}_{t}^{[\theta]},\mathcal{W}_{0:M-1};\theta) \notag\\
&\geq&0, \notag
\end{IEEEeqnarray}
where $(a)$ is because the files are independent of  the desired file index and the query, i.e., $I(\mathcal{W}_{0:M-1};\theta|\mathcal{Q}_{t}^{[\theta]})=0$, and the answer is a determined function of the received query and the files by \eqref{answer function}, i.e., $I(\mathbf{A}_{t}^{[\theta]};\theta|\mathcal{Q}_{t}^{[\theta]},\mathcal{W}_{0:M-1})=0$.
Hence,
\begin{IEEEeqnarray}{rCl}
&0=&I(\mathcal{Q}_{t}^{[\theta]},\mathbf{A}_{t}^{[\theta]},\mathcal{W}_{0:M-1};\theta) \notag\\
&\geq&I(\mathcal{Q}_{t}^{[\theta]},\mathbf{A}_{t}^{[\theta]},\mathcal{W}_{\Lambda};\theta) \notag\\
&\geq0&, \notag
\end{IEEEeqnarray}
which tell us $(\mathcal{Q}_{t}^{[\theta]},\mathbf{A}_{t}^{[\theta]},\mathcal{W}_{\Lambda})$ and $\theta$ are independent. Therefore, \eqref{Lemma:privacy} holds.
\end{IEEEproof}

\begin{Lemma}\label{Lemma appendix}
In the $(N,K,M)$ coded linear PIR scheme, for any $\theta\in[0:M),\Gamma\subseteq[0:N),\Lambda\subseteq[0:M)$ and given realization $\widetilde{\mathcal{Q}}_{0:N-1}^{[\theta]}$ of random queries ${\mathcal{Q}}_{0:N-1}^{[\theta]}$,
\begin{IEEEeqnarray}{rCl}
H(\mathbf{A}_{\Gamma}^{[\theta]}|\mathcal{W}_{\Lambda},\mathcal{Q}_{\Gamma}^{[\theta]})
&=&H(\mathbf{A}_{\Gamma}^{[\theta]}|\mathcal{W}_{\Lambda},\mathcal{Q}_{0:N-1}^{[\theta]}), \label{Lemma:random}\\
H(\mathbf{A}_{\Gamma}^{[\theta]}|\mathcal{W}_{\Lambda},\mathcal{Q}_{\Gamma}^{[\theta]}=\widetilde{\mathcal{Q}}_{\Gamma}^{[\theta]})
&=&H(\mathbf{A}_{\Gamma}^{[\theta]}|\mathcal{W}_{\Lambda},\mathcal{Q}_{0:N-1}^{[\theta]}=\widetilde{\mathcal{Q}}_{0:N-1}^{[\theta]}).
\label{Lemma:realization}
\end{IEEEeqnarray}
\end{Lemma}
\begin{IEEEproof}
As for \eqref{Lemma:random},
\begin{IEEEeqnarray}{lCl}
&&H(\mathbf{A}_{\Gamma}^{[\theta]}|\mathcal{W}_{\Lambda},\mathcal{Q}_{\Gamma}^{[\theta]})
-H(\mathbf{A}_{\Gamma}^{[\theta]}|\mathcal{W}_{\Lambda},\mathcal{Q}_{0:N-1}^{[\theta]}) \notag\\
&=&I(\mathbf{A}_{\Gamma}^{[\theta]};\mathcal{Q}_{[0:N)\setminus\Gamma}^{[\theta]}|\mathcal{W}_{\Lambda},\mathcal{Q}_{\Gamma}^{[\theta]}) \notag\\
&\leq&I(\mathbf{A}_{\Gamma}^{[\theta]},\mathcal{W}_{[0:M)\setminus\Lambda};\mathcal{Q}_{[0:N)\setminus\Gamma}^{[\theta]}|\mathcal{W}_{\Lambda},\mathcal{Q}_{\Gamma}^{[\theta]}) \notag\\
&=&I(\mathcal{W}_{[0:M)\setminus\Lambda};\mathcal{Q}_{[0:N)\setminus\Gamma}^{[\theta]}|\mathcal{W}_{\Lambda},\mathcal{Q}_{\Gamma}^{[\theta]})+
I(\mathbf{A}_{\Gamma}^{[\theta]};\mathcal{Q}_{[0:N)\setminus\Gamma}^{[\theta]}|\mathcal{W}_{0:M-1},\mathcal{Q}_{\Gamma}^{[\theta]}) \notag\\
&\overset{(a)}{=}&I(\mathcal{W}_{[0:M)\setminus\Lambda};\mathcal{Q}_{[0:N)\setminus\Gamma}^{[\theta]}|\mathcal{W}_{\Lambda},\mathcal{Q}_{\Gamma}^{[\theta]}) \notag\\
&\overset{(b)}{=}&0, \notag
\end{IEEEeqnarray}
where $(a)$ follows because $\mathbf{A}_{\Gamma}^{[\theta]}$ is only determined by $\mathcal{W}_{0:M-1}$ and  $\mathcal{Q}_{\Gamma}^{[\theta]}$ in \eqref{answer function}, i.e., $I(\mathbf{A}_{\Gamma}^{[\theta]};\mathcal{Q}_{[0:N)\setminus\Gamma}^{[\theta]}|\mathcal{W}_{0:M-1},\mathcal{Q}_{\Gamma}^{[\theta]})=0$;
$(b)$ is due to  the fact that the queries are independent of the files by \eqref{query independence} such that
\begin{IEEEeqnarray}{lCl}
0&=&I(\mathcal{W}_{0:M-1};\mathcal{Q}_{0:N-1}^{[\theta]}) \notag\\
&=&I(\mathcal{W}_{0:M-1};\mathcal{Q}_{\Gamma}^{[\theta]})+I(\mathcal{W}_{0:M-1};\mathcal{Q}_{[0:N)\setminus\Gamma}^{[\theta]}|\mathcal{Q}_{\Gamma}^{[\theta]}) \notag\\
&=&I(\mathcal{W}_{0:M-1};\mathcal{Q}_{\Gamma}^{[\theta]})+I(\mathcal{W}_{\Lambda};\mathcal{Q}_{[0:N)\setminus\Gamma}^{[\theta]}|\mathcal{Q}_{\Gamma}^{[\theta]})
+I(\mathcal{W}_{[0:M)\setminus\Lambda};\mathcal{Q}_{[0:N)\setminus\Gamma}^{[\theta]}|\mathcal{W}_{\Lambda},\mathcal{Q}_{\Gamma}^{[\theta]}) \notag\\
&\geq&I(\mathcal{W}_{[0:M)\setminus\Lambda};\mathcal{Q}_{[0:N)\setminus\Gamma}^{[\theta]}|\mathcal{W}_{\Lambda},\mathcal{Q}_{\Gamma}^{[\theta]}) \notag\\
&\geq&0. \notag
\end{IEEEeqnarray}

For \eqref{Lemma:realization}, we have
\begin{IEEEeqnarray}{lCl}
&&H(\mathbf{A}_{\Gamma}^{[\theta]}|\mathcal{W}_{\Lambda},\mathcal{Q}_{0:N-1}^{[\theta]}=\widetilde{\mathcal{Q}}_{0:N-1}^{[\theta]})\notag\\
&=&H(\mathcal{A}_{t,[0:M)}^{[\theta]}\mathbf{y}_t:t\in\Gamma|\mathcal{W}_{\Lambda},\mathcal{Q}_{0:N-1}^{[\theta]}=\widetilde{\mathcal{Q}}_{0:N-1}^{[\theta]})\notag\\
&=&H(\widetilde{\mathcal{A}}_{t,[0:M)}^{[\theta]}\mathbf{y}_t:t\in\Gamma|\mathcal{W}_{\Lambda})\notag\\
&=& H(\mathbf{A}_{\Gamma}^{[\theta]}|\mathcal{W}_{\Lambda},\mathcal{Q}_{\Gamma}^{[\theta]}=\widetilde{\mathcal{Q}}_{\Gamma}^{[\theta]}), \notag
\end{IEEEeqnarray}
where we use \eqref{linear answer function} and the fact that the answer matrices
${\mathcal{A}}_{t,[0:M)}^{[\theta]}$ ($t\in\Gamma$) are completely determined to be $\widetilde{\mathcal{A}}_{t,[0:M)}^{[\theta]}$ ($t\in\Gamma$) by  the corresponding query  realization
$\widetilde{\mathcal{Q}}_{\Gamma}^{[\theta]}$.
\end{IEEEproof}

\subsection*{Proof of Lemma \ref{answer independence}}

The proof of \eqref{Lem:answer indenp} can be found in \cite[Lemma 1]{Ulukus}.

In fact, for any $\Gamma\subseteq[0:N),|\Gamma|=K,\Lambda\subseteq[0:M)$, \eqref{Lem:answer indenp} can be equivalently written as
\begin{IEEEeqnarray}{lCl}
0&=&\sum\limits_{t\in\Gamma}H(\mathbf{A}_{t}^{[\theta]}|\mathcal{W}_{\Lambda},\mathcal{Q}_{t}^{[\theta]})-H(\mathbf{A}_{\Gamma}^{[\theta]}|\mathcal{W}_{\Lambda},\mathcal{Q}_{\Gamma}^{[\theta]}) \notag\\
&=&\sum\limits_{\widetilde{\mathcal{Q}}_{\Gamma}^{[\theta]}}\Pr(\mathcal{Q}_{\Gamma}^{[\theta]}=\widetilde{\mathcal{Q}}_{\Gamma}^{[\theta]})
\left[\sum\limits_{t\in\Gamma}H(\mathbf{A}_{t}^{[\theta]}|\mathcal{W}_{\Lambda},\mathcal{Q}_{t}^{[\theta]}=\widetilde{\mathcal{Q}}_{t}^{[\theta]})-
H(\mathbf{A}_{\Gamma}^{[\theta]}|\mathcal{W}_{\Lambda},\mathcal{Q}_{\Gamma}^{[\theta]}=\widetilde{\mathcal{Q}}_{\Gamma}^{[\theta]})\right].\label{Eqn_AQT}
\end{IEEEeqnarray}
While for every query realization $\widetilde{\mathcal{Q}}_{0:N-1}^{[\theta]}$, we always have
\begin{IEEEeqnarray*}{lCl}
&&\sum\limits_{t\in\Gamma}H(\mathbf{A}_{t}^{[\theta]}|\mathcal{W}_{\Lambda},\mathcal{Q}_{t}^{[\theta]}=\widetilde{\mathcal{Q}}_{t}^{[\theta]}) \notag\\
&\overset{(a)}{=}&\sum\limits_{t\in\Gamma}H(\mathbf{A}_{t}^{[\theta]}|\mathcal{W}_{\Lambda},\mathcal{Q}_{0:N-1}^{[\theta]}=\widetilde{\mathcal{Q}}_{0:N-1}^{[\theta]}) \notag\\
&\geq&H(\mathbf{A}_{\Gamma}^{[\theta]}|\mathcal{W}_{\Lambda},\mathcal{Q}_{0:N-1}^{[\theta]}=\widetilde{\mathcal{Q}}_{0:N-1}^{[\theta]}) \notag \\
&\overset{(b)}{=}&H(\mathbf{A}_{\Gamma}^{[\theta]}|\mathcal{W}_{\Lambda},\mathcal{Q}_{\Gamma}^{[\theta]}=\widetilde{\mathcal{Q}}_{\Gamma}^{[\theta]}),
\end{IEEEeqnarray*}
where $(a)$ and (b) follow from  applying \eqref{Lemma:realization} to the set $\{t\}$ and $\Gamma$ respectively.

That is, the terms in square bracket of  \eqref{Eqn_AQT}  are nonnegative. Therefore, they have to be  zero for all the query realizations $\widetilde{\mathcal{Q}}_{0:N-1}^{[\theta]}$ with positive probability, i.e., the necessary condition \textbf{P1} must be satisfied.
$\hfill\blacksquare$

\subsection*{Proof of Lemma \ref{recursion lemma}}
In fact,
\begin{IEEEeqnarray}{rCl}
H(\mathbf{A}_{0:N-1}^{[\theta]}|\mathcal{W}_{\Lambda},\mathcal{Q}_{0:N-1}^{[\theta]})&\overset{(a)}{\geq}&\frac{1}{\tbinom{N}{K}}\sum\limits_{\Gamma\subseteq[0:N):|\Gamma|=K}H(\mathbf{A}_{\Gamma}^{[\theta]}|\mathcal{W}_{\Lambda},\mathcal{Q}_{0:N-1}^{[\theta]})\notag\\
&\overset{(b)}{=}&\frac{1}{\tbinom{N}{K}}\sum\limits_{\Gamma\subseteq[0:N):|\Gamma|=K}H(\mathbf{A}_{\Gamma}^{[\theta]}|\mathcal{W}_{\Lambda},\mathcal{Q}_{\Gamma}^{[\theta]})\notag\\
&\overset{(c)}{=}&\frac{1}{\tbinom{N}{K}}\sum\limits_{\Gamma\subseteq[0:N):|\Gamma|=K}\sum\limits_{t\in\Gamma}H(\mathbf{A}_{t}^{[\theta]}|\mathcal{W}_{\Lambda},\mathcal{Q}_{t}^{[\theta]})\notag\\
&\overset{(d)}{=}&\frac{1}{\tbinom{N}{K}}\sum\limits_{\Gamma\subseteq[0:N):|\Gamma|=K}\sum\limits_{t\in\Gamma}H(\mathbf{A}_{t}^{[\theta']}|\mathcal{W}_{\Lambda},\mathcal{Q}_{t}^{[\theta']})\notag\\
&\overset{(e)}{=}&\frac{1}{\tbinom{N}{K}}\sum\limits_{\Gamma\subseteq[0:N):|\Gamma|=K}H(\mathbf{A}_{\Gamma}^{[\theta']}|\mathcal{W}_{\Lambda},\mathcal{Q}_{\Gamma}^{[\theta']})\notag\\
&\overset{(f)}{=}&\frac{1}{\tbinom{N}{K}}\sum\limits_{\Gamma\subseteq[0:N):|\Gamma|=K}H(\mathbf{A}_{\Gamma}^{[\theta']}|\mathcal{W}_{\Lambda},\mathcal{Q}_{0:N-1}^{[\theta']})\notag\\
&\overset{(g)}{\geq}&\frac{K}{N}H(\mathbf{A}_{0:N-1}^{[\theta']}|\mathcal{W}_{\Lambda},\mathcal{Q}_{0:N-1}^{[\theta']})\notag\\
&=&\frac{K}{N}\left( H(\mathbf{A}_{0:N-1}^{[\theta']},\mathcal{W}_{\theta'}|\mathcal{W}_{\Lambda},\mathcal{Q}_{0:N-1}^{[\theta']})-H(\mathcal{W}_{\theta'}|\mathcal{W}_{\Lambda},\mathbf{A}_{0:N-1}^{[\theta']},\mathcal{Q}_{0:N-1}^{[\theta']}) \right)\notag\\
&\overset{(h)}{=}&\frac{K}{N}\left(  H(\mathcal{W}_{\theta'}|\mathcal{W}_{\Lambda},\mathcal{Q}_{0:N-1}^{[\theta']})+H(\mathbf{A}_{0:N-1}^{[\theta']}|\mathcal{W}_{\Lambda},\mathcal{W}_{\theta'},\mathcal{Q}_{0:N-1}^{[\theta']}) \right)\notag\\
&\overset{(i)}{=}&\frac{KL}{N}+\frac{K}{N}H(\mathbf{A}_{0:N-1}^{[\theta']}|\mathcal{W}_{\Lambda},\mathcal{W}_{\theta'},\mathcal{Q}_{0:N-1}^{[\theta']}), \notag
\end{IEEEeqnarray}
where $(a)$ is because
\begin{eqnarray}\label{interference mutual deter}
H(\mathbf{A}_{0:N-1}^{[\theta]}|\mathcal{W}_{\Lambda},\mathcal{Q}_{0:N-1}^{[\theta]})\overset{(j)}{\geq} H(\mathbf{A}_{\Gamma}^{[\theta]}|\mathcal{W}_{\Lambda},\mathcal{Q}_{0:N-1}^{[\theta]}), \,\Gamma\subseteq[0:N), ~|\Gamma|=K;
\end{eqnarray}
$(b)$ and $(f)$ are due to  \eqref{Lemma:random};
$(c)$ and $(e)$ follow from \eqref{Lem:answer indenp};
$(d)$ is because of \eqref{Lemma:privacy};
$(g)$ is due to the well-known Han's inequality \cite[Theorem 17.6.1]{Element of IT}:
\begin{equation}
\frac{1}{K\tbinom{N}{K}}\sum\limits_{\Gamma\subseteq[0:N):|\Gamma|=K}H(\mathbf{A}_{\Gamma}^{[\theta]}|\mathcal{W}_{\Lambda},\mathcal{Q}_{0:N-1}^{[\theta]})\geq\frac{1}{N}H(\mathbf{A}_{0:N-1}^{[\theta]}|\mathcal{W}_{\Lambda},\mathcal{Q}_{0:N-1}^{[\theta]})\notag
\end{equation}
for any set $\Lambda\subseteq[0:M)$; $(h)$ follows from the correctness constraint in \eqref{correctness constrain} such that $\mathbf{A}_{0:N-1}^{[\theta']}$ and $\mathcal{Q}_{0:N-1}^{[\theta']}$ can decode the requested file $\mathcal{W}_{\theta'}$, i.e., $H(\mathcal{W}_{\theta'}|\mathcal{W}_{\Lambda},\mathbf{A}_{0:N-1}^{[\theta']},\mathcal{Q}_{0:N-1}^{[\theta']})=0$; $(i)$ is due to \eqref{query independence} where queries are independent of the files such that $H(\mathcal{W}_{\theta'}|\mathcal{W}_{\Lambda},\mathcal{Q}_{0:N-1}^{[\theta']})=H(\mathcal{W}_{\theta'})=L$ by
\eqref{Eqn_W_Size}.

If the equality in $(a)$ holds, then the  equality in $(j)$ of \eqref{interference mutual deter} must hold, i.e., for every $\Lambda\subseteq[0:M),\Gamma\subseteq[0:N),|\Gamma|=K$, and $\theta\in\Lambda$,
\begin{IEEEeqnarray}{rCl}
0&=&H(\mathbf{A}_{0:N-1}^{[\theta]}|\mathcal{W}_{\Lambda},\mathcal{Q}_{0:N-1}^{[\theta]})-H(\mathbf{A}_{\Gamma}^{[\theta]}|\mathcal{W}_{\Lambda},\mathcal{Q}_{0:N-1}^{[\theta]})\notag\\
&=& \sum\limits_{\widetilde{\mathcal{Q}}_{0:N-1}^{[\theta]}}\Pr(\mathcal{Q}_{0:N-1}^{[\theta]}=\widetilde{\mathcal{Q}}_{0:N-1}^{[\theta]})\left[H(\mathbf{A}_{0:N-1}^{[\theta]}|\mathcal{W}_{\Lambda},\mathcal{Q}_{0:N-1}^{[\theta]}=\widetilde{\mathcal{Q}}_{0:N-1}^{[\theta]})-
H(\mathbf{A}_{\Gamma}^{[\theta]}|\mathcal{W}_{\Lambda},\mathcal{Q}_{0:N-1}^{[\theta]}=\widetilde{\mathcal{Q}}_{0:N-1}^{[\theta]})\right].
\label{kk}
\end{IEEEeqnarray}
Whereas for every query realization $\widetilde{\mathcal{Q}}_{0:N-1}^{[\theta]}$ with positive probability,
\begin{IEEEeqnarray*}{rCl}
H(\mathbf{A}_{0:N-1}^{[\theta]}|\mathcal{W}_{\Lambda},\mathcal{Q}_{0:N-1}^{[\theta]}=\widetilde{\mathcal{Q}}_{0:N-1}^{[\theta]})&\ge &H(\mathbf{A}_{\Gamma}^{[\theta]}|\mathcal{W}_{\Lambda},\mathcal{Q}_{0:N-1}^{[\theta]}=\widetilde{\mathcal{Q}}_{0:N-1}^{[\theta]})\notag\\
&\overset{(k)}{=}&H(\mathbf{A}_{\Gamma}^{[\theta]}|\mathcal{W}_{\Lambda},\mathcal{Q}_{\Gamma}^{[\theta]}=\widetilde{\mathcal{Q}}_{\Gamma}^{[\theta]}),
\end{IEEEeqnarray*}
where $(k)$ is due to \eqref{Lemma:realization}.

This is to say,    the terms in square bracket of  \eqref{kk}  are nonnegative.
Consequently, to make the equality in $(a)$ hold,
they have to be  zero for all the query realizations $\widetilde{\mathcal{Q}}_{0:N-1}^{[\theta]}$ with positive probability, i.e., the necessary condition \textbf{P2} must be satisfied.
$\hfill\blacksquare$

\subsection*{Proof of Lemma \ref{intial step}}
Notice that,
\begin{IEEEeqnarray}{rCl}
H(\mathbf{A}_{0:N-1}^{[\theta]}|\mathcal{W}_{\theta},\mathcal{Q}_{0:N-1}^{[\theta]})&=&H(\mathbf{A}_{0:N-1}^{[\theta]}|\mathcal{Q}_{0:N-1}^{[\theta]})-I(\mathbf{A}_{0:N-1}^{[\theta]};\mathcal{W}_{\theta}|\mathcal{Q}_{0:N-1}^{[\theta]})\notag\\
&=&H(\mathbf{A}_{0:N-1}^{[\theta]}|\mathcal{Q}_{0:N-1}^{[\theta]})-H(\mathcal{W}_{\theta}|\mathcal{Q}_{0:N-1}^{[\theta]})+H(\mathcal{W}_{\theta}|\mathbf{A}_{0:N-1}^{[\theta]},\mathcal{Q}_{0:N-1}^{[\theta]})\notag\\
&\overset{(a)}{=}&H(\mathbf{A}_{0:N-1}^{[\theta]}|\mathcal{Q}_{0:N-1}^{[\theta]})-L\notag\\
&\overset{(b)}{\leq}&\sum\limits_{t=0}^{N-1}H(\mathbf{A}_{t}^{[\theta]}|\mathcal{Q}_{0:N-1}^{[\theta]})-L\notag\\
&\overset{(c)}{=}&\sum\limits_{t=0}^{N-1}H(\mathbf{A}_{t}^{[\theta]}|\mathcal{Q}_{t}^{[\theta]})-L\notag\\
&=&\sum\limits_{t=0}^{N-1}\sum\limits_{\widetilde{\mathcal{Q}}_{t}^{[\theta]}}\Pr(\mathcal{Q}_{t}^{[\theta]}=\widetilde{\mathcal{Q}}_{t}^{[\theta]})H(\mathbf{A}_{t}^{[\theta]}|\mathcal{Q}_{t}^{[\theta]}=\widetilde{\mathcal{Q}}_{t}^{[\theta]})-L\notag\\
&\overset{(d)}{=}&\sum\limits_{t=0}^{N-1}\sum\limits_{\widetilde{\mathcal{Q}}_{t}^{[\theta]}}\Pr(\mathcal{Q}_{t}^{[\theta]}=\widetilde{\mathcal{Q}}_{t}^{[\theta]})H(\widetilde{\mathcal{A}}_{t,[0:M)}^{[\theta]}\mathbf{y}_t)-L\notag\\
&\overset{(e)}{\leq}&\sum\limits_{t=0}^{N-1}\sum\limits_{\widetilde{\mathcal{Q}}_{t}^{[\theta]}}\Pr(\mathcal{Q}_{t}^{[\theta]}=\widetilde{\mathcal{Q}}_{t}^{[\theta]})\ell_t-L \notag\\
&=&\sum\limits_{t=0}^{N-1}\mathbb{E}\left[ \ell_t\right]-L \notag\\
&=&\frac{L}{R}-L, \notag
\end{IEEEeqnarray}
where $(a)$ follows from the correctness constraint in \eqref{correctness constrain} and the fact that queries are independent from the files, i.e., $H(\mathcal{W}_{\theta}|\mathcal{Q}_{0:N-1}^{[\theta]})=H(\mathcal{W}_{\theta})=L$; $(b)$ is due to the inequality of joint entropy; $(c)$ follows from \eqref{Lemma:random} by setting $\Gamma=\{t\}$ and $\Lambda=\emptyset$; $(d)$ is by the definition of the linear coded PIR in \eqref{linear answer function} and then $\mathbf{A}_{t}^{[\theta]}=\mathcal{A}_{t,[0:M)}^{[\theta]}\mathbf{y}_t=\widetilde{\mathcal{A}}_{t,[0:M)}^{[\theta]}\mathbf{y}_t$ for  the received  query  realization $\widetilde{\mathcal{Q}}_{t}^{[\theta]}$; $(e)$ is due to the principle of maximum entropy, i.e., $H(\widetilde{\mathcal{A}}_{t,[0:M)}^{[\theta]}\mathbf{y}_t)\leq \ell_t$, where $\ell_t$ is the length of $\widetilde{\mathcal{A}}_{t,[0:M)}^{[\theta]}\mathbf{y}_t$.


In fact, the equality in $(b)$ holds if and only if
\begin{IEEEeqnarray}{rCl}
0&=&\sum\limits_{t=0}^{N-1}H(\mathbf{A}_{t}^{[\theta]}|\mathcal{Q}_{0:N-1}^{[\theta]})-H(\mathbf{A}_{0:N-1}^{[\theta]}|\mathcal{Q}_{0:N-1}^{[\theta]})\notag\\
&=&\sum\limits_{\widetilde{\mathcal{Q}}_{0:N-1}^{[\theta]}}\Pr(\mathcal{Q}_{0:N-1}^{[\theta]}=\widetilde{\mathcal{Q}}_{0:N-1}^{[\theta]})\left[\sum\limits_{t=0}^{N-1}H(\mathbf{A}_{t}^{[\theta]}|\mathcal{Q}_{0:N-1}^{[\theta]}=\widetilde{\mathcal{Q}}_{0:N-1}^{[\theta]})
-H(\mathbf{A}_{0:N-1}^{[\theta]}|\mathcal{Q}_{0:N-1}^{[\theta]}=\widetilde{\mathcal{Q}}_{0:N-1}^{[\theta]})\right].\label{Eqn_AQH}
\end{IEEEeqnarray}
While
\begin{IEEEeqnarray*}{rCl}
\sum\limits_{t=0}^{N-1}H(\mathbf{A}_{t}^{[\theta]}|\mathcal{Q}_{t}^{[\theta]}=\widetilde{\mathcal{Q}}_{t}^{[\theta]})&\overset{(f)}{=}&
\sum\limits_{t=0}^{N-1}H(\mathbf{A}_{t}^{[\theta]}|\mathcal{Q}_{0:N-1}^{[\theta]}=\widetilde{\mathcal{Q}}_{0:N-1}^{[\theta]})\notag\\
&\ge &H(\mathbf{A}_{0:N-1}^{[\theta]}|\mathcal{Q}_{0:N-1}^{[\theta]}=\widetilde{\mathcal{Q}}_{0:N-1}^{[\theta]}),
\end{IEEEeqnarray*}
where $(f)$ follows from \eqref{Lemma:realization} by setting $\Gamma=\{t\}$ and $\Lambda=\emptyset$.

This means that the terms in square bracket of  \eqref{Eqn_AQH} are nonnegative. Hence, to make the equality in $(b)$ hold,
they have to be  zero for all the query realizations $\widetilde{\mathcal{Q}}_{0:N-1}^{[\theta]}$ with positive probability, i.e., the necessary condition \textbf{P3} must be satisfied.
$\hfill\blacksquare$

\end{document}